\documentclass[twocolumn, numberedappendix]{aastex63}

\usepackage{amstext}
\usepackage{amsmath}
\usepackage{amssymb}
\usepackage{graphicx}
\usepackage{enumitem}

\newcommand{\beq}{\begin{equation}}
\newcommand{\eeq}{\end{equation}}
\newcommand{\beqar}{\begin{align}}
\newcommand{\eeqar}{\end{align}}

\newcommand{\Md}{M_{\rm d}}
\newcommand{\Rd}{R_{\rm d}}
\newcommand{\Ma}{M_{\rm a}}
\newcommand{\gacc}{\gamma_{\rm a}}
\newcommand{\gloss}{\gamma_{\rm loss}}
\newcommand{\gd}{\gamma_{\rm d}}
\newcommand{\gL}{\gamma_{L_2}}
\newcommand{\lacc}{l_{\rm a}}

\newcommand{\ld}{l_{\rm d}}
\newcommand{\lL}{l_{L_2}}

\shorttitle{}

\shortauthors{}

\begin{document}

\title{Runaway Coalescence of Pre-Common-Envelope Stellar Binaries}

\author[0000-0002-1417-8024]{Morgan MacLeod}
\affiliation{Harvard-Smithsonian Center for Astrophysics, 60 Garden Street, Cambridge, MA, 02138, USA}
\email{morgan.macleod@cfa.harvard.edu}

\author[0000-0003-4330-287X]{Abraham Loeb}
\affiliation{Harvard-Smithsonian Center for Astrophysics, 60 Garden Street, Cambridge, MA, 02138, USA}

\begin{abstract}
We study the process of runaway, unstable Roche lobe overflow in coalescing binary systems and its dependence on the properties of the binary involved. We create three-dimensional hydrodynamic models of binary coalescences, and follow them through a phase of increasing Roche lobe overflow until the accretor is engulfed by the donor at the onset of a common envelope phase. In these models, we vary binary properties of mass ratio, donor structure and spin, and equation of state through the gas adiabatic index. We compare the numerical results to semi-analytic models of binary orbit evolution based on mass and angular momentum exchange between two point masses. Using our hydrodynamic simulations, we measure the key parameters: the donor mass loss rate and the angular momentum exchanged per unit mass loss from the donor. Using these calibrations, the semi-analytic model closely reproduces the escalating mass loss and runaway orbital decay observed in the hydrodynamic models. The semi-analytic model accurately reproduces the major differences in orbit evolution that arise with varying mass ratio and donor structure.  We encapsulate the semi-analytic model in a publicly-released python package {\tt RLOF}. We apply this model to the observed period decay and subsequent merger of the binary V1309 Sco, and find that it can simultaneously reproduce the observed orbital decay and time of outburst. We further demonstrate that there is a relationship between period derivative and second derivative that can be a useful metric for evaluating candidate merging binaries.
\end{abstract}


\section{Introduction}

Binary and multiple star systems are common across the stellar mass spectrum and across stellar evolutionary type \citep{2013ARA&A..51..269D}. They exist with relatively equal frequency in configurations that range from extremely compact to so widely separated as to be barely gravitationally bound \citep[e.g.][]{2017ApJS..230...15M}. The closest of these systems directly interact as the stars evolve and change in radius \citep{2017PASA...34....1D}. These interactions have many possible outcomes, among them the runaway coalescence of the two stars following a phase of unstable mass loss. These runaway coalescence events mark the onset of common envelope phases, in which one star engulfs its more compact companion and the two stellar cores spiral closer within a shared envelope \citep{1976IAUS...73...75P}. 

In at least one case, such a runaway orbital decay has been directly observed -- in the pre-outburst lightcurve of V1309 Sco \citep{2011A&A...528A.114T}. This binary system was identified as an eclipsing binary with a decreasing orbital period in archival data from the years prior to its 2008 outburst \citep{2010A&A...516A.108M,2011A&A...528A.114T}.  \citet{2014ApJ...788...22P} and \citet{2017ApJ...850...59P} have demonstrated that a model that reproduces the orbital decay and lightcurve morphology in this object involves non-conservative mass loss, in which material trails away from the binary near the outer Lagrange point, $L_2$ \citep[also see][]{2016MNRAS.455.4351P,2016MNRAS.461.2527P,2017MNRAS.471.3200M}. 

In other cases, orbital decay is inferred to precede outbursts similar to V1309 Sco, which define the emerging class of luminous red novae \citep[e.g.][]{2014MNRAS.443.1319K,2015ATel.7173....1D,2015A&A...578L..10K,2015ApJ...805L..18W,2017ApJ...834..107B,2017ApJ...835..282M, 2018PASP..130c4202A,2019A&A...630A..75P}. Recurring features of these systems include precursor activity  -- slow brightening or dimming -- prior to outburst \citep{2011A&A...528A.114T,2015ATel.7173....1D,2017ApJ...834..107B,2017ApJ...850...59P}, and a very rapid timescale of days to weeks rise to peak of the flares themselves \citep{2010A&A...516A.108M,2017ApJ...835..282M}. A priori, the origin of this rapid timescale in a binary system with similar orbital period is not immediately obvious. \citet{2018ApJ...863....5M} studied a  hydrodynamic simulation to demonstrate that these rapid timescales originate in the runaway of orbital decay and mass ejection that result as the end stage of unstable mass loss. \citet{2018ApJ...863....5M} also found that a semi-analytic model of mass and angular momentum exchange between point masses in a binary pair reproduced the main features of their simulated model. 

A complicating property of binary and multiple star interactions is that binaries come in many combinations of orbital configuration and constituent objects. In this paper, we present a suite of hydrodynamic models of binary coalescence from separations near the Roche limit until the more-compact accretor is engulfed within the donor. We vary properties of binary mass ratio, donor star structure and spin relative to the orbit, and gas equation of state. We study the effects of these differences on the dynamics of runaway orbital decay and the binary coalescence that leads to a common envelope phase. In so doing, we measure the key parameters that define the point-mass orbit evolution model. We describe and release a python software package for rapid orbit integration using these parameters as calibrated by our hydrodynamic simulations.

The outline of this paper is as follows. In Section \ref{sec:analytic}, we describe the analytic framework of mass loss in a point-mass binary pair, and the key unknown quantities in this model. In Section \ref{sec:method}, we describe our hydrodynamic simulation method and model parameters. In Section \ref{sec:donormdot}, we discuss the measurement of the mass loss rate from the donor star. Section \ref{sec:gloss} discusses the measurement of the specific angular momentum of material lost from the binary. Section \ref{sec:orb} discusses the resultant orbital evolution, and combines these measurements to produce reconstructed orbital evolutions from analytic theory. In Section \ref{sec:discussion}, we discuss some implications of our findings for observed systems like V1309 Sco, and  Section \ref{sec:conclusion} summarizes our conclusions.

\section{Analytic Framework}\label{sec:analytic}
We start by reviewing the semi-analytic framework for binary orbital evolution and identify key, unkown parameters that we will measure in our simulated systems \citep[see the review of][for a more extended discussion]{1981ARA&A..19..277S}. 

\subsection{Orbital Evolution}
The coupling of orbital evolution to mass and angular momentum exchange in a binary system can be expressed as an ordinary differential equation in the limit where we treat the binary components as point masses in circular orbit. 
Then, 
\begin{align}
{\dot a \over a} =  -2  \frac{\dot \Md}{\Md} 
 \left[ 1- \beta \frac{\Md}{\Ma}   - \left(1-\beta \right)\left( \gloss + {1\over 2}\right) {\Md \over M}  \right], 
\end{align}
Here $\Md$ is the mass of the donor star, $\Ma$ is the mass of the accretor star, and  $M=\Md + \Ma$. The binary separation of the circular orbit is $a$, and $\gamma_{\rm loss}$ is a dimensionless specific angular momentum of material leaving the binary, 
\beq
\gamma_{\rm loss} = {l_{\rm loss}  \over l_{\rm bin}},
\eeq
 the ratio of the specific angular momentum, $l_{\rm loss}$, of lost material to the specific angular momentum of the binary, $l_{\rm bin} = \Md \Ma / M^2 \sqrt{GMa}$. 
Finally, $\beta$ represents a fraction of mass lost from the donor that is captured by the accretor, $\dot \Ma = - \beta \dot \Md$.  
In what follows, we will consider the limit $\beta \rightarrow 0$ of the expression above -- because this is the case that has relevance for comparison to our hydrodynamic simulations --  such that, 
\beq\label{orbit_analytic}
{\dot a \over a}  = -2 {\dot \Md  \over \Md } \left[ 1  - \left(\gamma_{\rm loss}+{1\over2}\right) {\Md \over M}\right] .
\eeq
This represents the case of fully non-conservative mass loss from a binary system \citep{1963ApJ...138..471H}.\footnote{Although similar expressions are widely used, a particularly pedagogical introduction is given in  O. Pols' binary-evolution notes, Chapter 7, online at \url{http://www.astro.ru.nl/~onnop/education/binaries_utrecht_notes/} }

 It is apparent that the rate of orbital evolution described by equation \eqref{orbit_analytic} is a direct result of the uncertain parameters of $\dot \Md$ and $\gloss$.  With knowledge of these parameters, one may estimate the orbit evolution rate (for example the number of orbits remaining until a binary coalesces) as a function of binary properties. 

\subsection{The Donor Mass Loss Rate}

The strongest influence on  the  mass loss rate stems from the binary separation; in binary systems where the donor star overflows its Roche lobe more, the  mass loss rate should be higher. However, many of the donor stars properties (including its structure and rotation) may also affect the mass loss rate. Measuring those effects is one of the goals of this paper. To do so, we will use a baseline prediction from \citet{1972AcA....22...73P}, who estimated the mass loss rate of a polytropic donor star of index $n$ to be,
\beq\label{mdot_analytic}
\dot \Md = - \alpha   {\Md \over \tau } \left( {R_{\rm d} - R_L \over R_{\rm d}} \right)^{n+{3\over 2}}.
\eeq
Here $\Md$ and $R_{\rm d}$ are the mass and radius of the donor, $R_L$ is the radius of the Roche lobe \citep{1983ApJ...268..368E}, and $\tau$ is the binary orbital period, $n=1/(\Gamma_{\rm s}-1)$, and $\Gamma_{\rm s}$ is the polytropic index (the logarithmic derivative of gas pressure with respect to density). 

Some numerical confirmation of this approximation has come through a study by \citet{1987MNRAS.229..383E}, using two-dimensional hydrodynamic simulations.  \citet{2018ApJ...863....5M} applied this expression to one simulation of binary coalescence, and found that $\alpha \approx 1$ reproduced the model behavior quite well (their Figure 7). 

Improvements on this model have been presented in the literature with a focus on low  mass loss rates, and the behavior of material near the optical photosphere. For example, \citet{2017ApJ...835..145J} presented a detailed model that accounts for the presence of atmospheric material above the optical photosphere radius -- which determines the degree of Roche lobe overflow in equation \eqref{mdot_analytic}.    \citet{2015MNRAS.449.4415P} demonstrated the importance of the thermodynamics of the outermost layers involved in the mass-transferring flow, especially in giant-star donors. Because the optical depth becomes very low near the donor surface,  \citet{2015MNRAS.449.4415P} show that there is always some material that can thermally adjust and this non-adiabatic behavior can impact the donor's mass loss rate as a function of degree of Roche lobe overflow, and, in turn, the stability of mass transfer. Non-adiabatic properties of donor-object's outer layers may also be important on objects with hot, irradiated surface layers, as in x-ray binaries or close in planets \citep{2017ApJ...835..145J}. Despite these important effects, we focus on the simple expression of equation \eqref{mdot_analytic} for comparison to our numerical simulations in Section \ref{sec:donormdot} because we  are only able to study phases of very deep Roche lobe overflow compared to most models of binary mass exchange and because the more detailed thermodynamics of radiatively cooling flows are not present in our hydrodynamic calculations.

\subsection{The Value of $\gamma_{\rm loss}$ }

A second key uncertainty that we will use our numerical models to asses is the specific angular momentum of material lost from the binary, parameterized by $\gamma_{\rm loss}$. Although a priori we do not know the value of this parameter in a given binary system, several guiding values serve as useful benchmarks  \citep{1963ApJ...138..471H}: 
\begin{enumerate}[label=(\roman*)]
\item The specific angular momentum of the donor star, $\ld = \Ma^2 / M^2 \sqrt{GMa}$, which implies $\gd=\Ma/\Md$;
\item The specific angular momentum of the accretor star, $\lacc = \Md^2 / M^2 \sqrt{GMa}$, which implies $\gacc=\Md/\Ma$; and
\item The specific angular momentum of the outer Lagrange point near the secondary, $L_2$,  $\lL\approx 1.2^2  \sqrt{GMa}$, which implies  $\gL \approx 1.2^2 M^2 / (\Md\Ma)$ \citep{1998CoSka..28..101P}. 
\end{enumerate}
For illustrative purposes, case (i) might occur if the donor were losing material via a wind that is unaffected by the accretor. Case (ii) might occur if the donor transferred material to the vicinity of the accretor, then it was expelled (this case is sometimes described as isotropic re-emission). Finally, the $L_2$ Lagrange point represents the lower of two saddle points in the effective potential of the binary system from which material could conceivably escape from the binary, leading to case (iii). 

Across the range of binary system behavior, we expect situations that are approximated by each of these example cases. In previous work, \citet{2018ApJ...863....5M} measured the $\gamma_{\rm loss}$ for a particular binary system with accretor to donor mass ratio of 0.3 and found that it lay mid-way between $\gacc$ and $\gL$.

\section{Method and Simulation Models}\label{sec:method}

Here we briefly summarize our numerical method, described in detail in \citet{2018ApJ...863....5M} and \citet{2018ApJ...868..136M}, and describe the models analyzed in this paper. 

\subsection{Numerical Method}

We model the coalescence of binary systems using the {\tt Athena++} hydrodynamics code (Stone, J. M. et al., in preparation).\footnote{version 1.0, url: https://princetonuniversity.github.io/athena}  To do so, we solve the inviscid hydrodynamic conservation equations of mass, momentum, and energy with additional source terms corresponding to the gravitational potential of the binary system. Our models are computed in the orbiting (but non-rotating) reference frame of the donor star in the binary system. We adopt an ideal gas equation of state with adiabatic index $\gamma_{\rm ad}$. A full description of the equations solved, source terms, and tests is given in \citet{2018ApJ...863....5M}. A summary of these methods is reported in \citet{2018ApJ...868..136M}.

In general, we model a giant-star donor that fills its Roche lobe and transfers mass onto a less-massive, more-compact donor (treated as a softened point mass).  
 These models correspond to the $\beta = 0$ case described in Section \ref{sec:analytic} because we do not allow a mechanism for mass to be absorbed from the mesh or added to the accretor point mass.
The initial mass of the donor is defined as $M_1$ and the mass of the accretor is $M_2$. The mass ratio of accretor to donor is $q \equiv M_2/M_1$, which is equivalent to $\Ma/\Md$ when the simulation is initialized. Our simulations are performed  in  dimensionless units such that the donor's original mass, radius, and the gravitational constant are all one, $M_1=R_1=G=1$. Therefore, the time unit in the simulation is the characteristic donor-star dynamical time, $t_{\rm dyn,1}=(R_1^3/G M_1)^{1/2}$. Simulated results may be rescaled to physical systems from these values. Several examples of this rescaling are given in Table 1 of \citet{2018ApJ...863....5M}. 

Following initialization in the hydrodynamic mesh, we gradually initialize the fully dynamical part of the problem solution as follows. The donor star is mapped from one-dimensional model onto the three-dimensional mesh, after which time it is ``relaxed" for 15$t_{\rm dyn,1}$, during which time, motion in the $r$ and $\theta$ directions of the spherical polar mesh is damped. Next, we progressively turn on the gravitational action of $M_2$ on the gas over the following 15$t_{\rm dyn,1}$. After 30$t_{\rm dyn,1}$, the model is fully active.

\subsection{Simulation Models}

\begin{table*}[tb]
\begin{center}
\hspace{-3cm}
\begin{tabular}{ccccccccccc}
Model & $M_1$ & $m_1$ & $M_2$ & $a_0$ & $\gamma_{\rm ad}$ & $\Gamma_{\rm s}$ & $f_{\rm corot}$ & $R_{V} $ & $\langle \alpha \rangle$  & $\langle \gloss \rangle $ \\
\hline
A &1&  0.41 & 0.1 & 1.73 & 5/3 & 5/3 & 1.0 & 1.045 & 0.714 & 12.55 \\
B &1& 0.41 & 0.1 & 1.73 & 5/3 & 5/3 & 0.67 & 1.022 & 0.818 & 13.69 \\
C &1& 0.41 & 0.1 & 1.73 & 5/3 & 5/3 & 0.33 & 1.009 & 0.971 & 14.42  \\
D &1& 0.41 & 0.1 & 1.73 & 5/3 & 5/3 & 0.0  & 1.007 & 1.164 & 14.22 \\
\hline
E &1& 0.41 & 0.03 & 1.51 & 5/3 & 5/3 & 1.0  & 1.065 & 0.255 & 21.70 \\
F &1& 0.41 & 0.3 & 2.06 & 5/3 & 5/3 & 1.0 & 1.033 & 1.211 & 6.20 \\
\hline
G &1& 0.68 & 0.1 & 1.55 & 1.35 & 1.35 & 1.0 & 1.060 & 0.453 & 15.87 \\
H &1& 0.68 & 0.1 & 1.55 & 1.5 & 1.35 & 1.0  & 1.044 & 0.737 & 18.27 \\
I &1& 0.68 & 0.1 & 1.55 & 5/3 & 1.35 & 1.0 & 1.035 & 1.081 & 19.28 \\
\hline
J &1& 0.68 & 0.3 & 1.75 & 1.35 & 1.35 & 1.0 & 1.039 & 0.765 & 7.58 \\
K &1& 0.68 & 0.3 & 1.75 & 1.5 & 1.35 & 1.0 & 1.026 & 1.486 & 8.68  \\
L &1& 0.68 & 0.3 & 1.75 & 5/3 & 1.35 & 1.0  & 1.021 & 2.640 & 8.46 \\
\hline
M & 1 & 0.41 & 0.1 & 1.55 & 5/3 & 5/3 & 1.0 & 1.046 & 0.715 & 11.56 \\
\hline 
\end{tabular}
\caption{Parameters of model binary systems simulated.  Columns are: donor initial mass, $M_1$, donor central mass, $r<0.3R_1$, $m_1$, accretor mass $M_2$, initial separation, $a_0$, adiabatic index, $\gamma_{\rm ad}$, donor polytropic structural index, $\Gamma_{\rm s}$, fractional spin synchronization $f_{\rm corot}$, volume averaged rotating donor radius, $R_V$, averaged  mass loss coefficient, $\langle \alpha \rangle$, averaged specific angular momentum of expelled material, $\langle \gloss \rangle$. Models A-D represent variations in initial spin synchronization. Models E and F represent varying mass ratio. Models G-I and Models J-L are variations $\gamma_{\rm ad}$ with $\Gamma_{\rm s}=1.35$, and $q=0.1$ and $q=0.3$, respectively. Finally, model M is identical to A, except that it starts at a separation consistent  with models G-I.  }
\label{simtable}
\end{center}
\end{table*}

We report on a range of simulated binary coalescences in this paper. All models are initialized at separations greater than $R_1$ at which the donor overflows its Roche lobe. This  mass loss draws the binary closer, leading to the eventual system coalescence. Our models stop when the binary separation reaches $0.6R_1$. With these common features we survey across binary mass ratio, donor-star synchronization state, and donor star structure. A summary of these models is reported in Table \ref{simtable}. 

Our models adopt shared computational domain and spatial resolution. Our spherical polar computational domain surrounds the donor star core and extends from $0.3R_1 \leq r \leq 100 R_1$. The angular coordinates cover the full $4\pi$ solid angle: $0\leq \theta \leq \pi$ and $0\leq \phi \leq 2\pi$.  The computational domain is spatially decomposed into mesh elements. The base level mesh features $384\times192\times384$ zones in $r$, $\theta$ and $\phi$. The radial zones are logarithmically spaced, while the angular zones are uniformly spaced (ensuring that zone shapes remain approximately cubical).\footnote{Near the poles, we apply a special treatement, described in \citet{2018ApJ...863....5M,2018ApJ...868..136M}, that merges zones to avoid the timestep restrictions of very long, narrow zone shapes. This code and examples are made public at \url{https://github.com/morganemacleod/athena-public-version/tree/polar-zone-avg}. } Two levels of static mesh refinement increase the spatial resolution within the donor star and in the binary equatorial plane (for $r<6R_1$). The softening radius around $M_2$ is $0.05 R_1$. 

Donor star structure in all cases is that of a polytrope featuring a non-zero core mass. Each is selected to have a core of $0.25M_1$ within $0.1R_1$.  We parameterize these models in terms of their structural polytropic index $\Gamma_{\rm s}$, such that $P \propto \rho^{\rm \Gamma_{s}}$ within the model. Stable solutions of hydrostatic equilibrium are possible where $\gamma_{\rm ad} \geq \Gamma_{\rm s}> 4/3$.  Upon initialization within the hydrodynamic method, the inner portion of the donor-star profile is excised by the inner radial boundary ($0.3R_1$ in this case) and is added to a point mass representing the enclosed donor material and core, $m_1$. Summaries of these properties are given in Table \ref{simtable}. 

Our models with $\Gamma_{\rm s} =\gamma_{\rm ad} = 5/3$ are most relevant to lower mass systems (donor mass less than approximately $8M_\odot$) with convective envelopes. In these stars, gas pressure dominates the equation of state of the stellar envelope. Our models with $\Gamma_{\rm s} = 1.35$ and $\gamma_{\rm ad}=5/3$ are relevant to lower mass, gas pressure dominated stellar envelopes that are radiative rather than convective (and therefore not isentropic). Radiative envelopes are present in main sequence stars and early evolved stars crossing from the main sequence to the Hayashi track in the Hertzsprung-Russell diagram. Finally, models with $\Gamma_{\rm s} = \gamma_{\rm ad} = 1.35$ are relevant to the convective, isentropic envelopes of high-mass stars (approximately those greater than $8M_\odot$) which have an equation of state dominated by radiation pressure. 

We examine different degrees of donor star rotation but fix the orientation to be parallel to the binary system's angular momentum vector. In each case, we initialize the donor in solid body rotation with frequency $\Omega_{\rm spin}$. We therefore parameterize our models with the factor $f_{\rm corot}$, which describes the relation of  $\Omega_{\rm spin}$ to the orbital frequency, $\Omega_{\rm spin} = f_{\rm corot} \Omega_{\rm orb}$. Corotating models, $f_{\rm corot}=1$, are of relevance to binaries in which tidal dissipation has acted to synchronize the donor star's rotation to the orbital motion. Situations in which systems may be in asynchronous rotation include when there is too little angular momentum in the orbit to fully spin up the donor (the \citet{1879RSPS...29..168D} tidal instability \citep{1980A&A....92..167H}) or situations where the donor or orbit is evolving more rapidly than tides can synchronize the donor's rotation to the orbital motion \citep[one such example is a hierarchical triple system undergoing secular oscillations, e.g. ][]{2014ApJ...793..137N}.

Models are initialized at a separation $a_0$, listed in Table \ref{simtable}. Compared to the $\Gamma_{\rm s}=5/3$ models, the $\Gamma_{\rm s}=1.35$ donor stars have lower-mass, more centrally concentrated envelopes. Additionally, as we will discuss in Section \ref{sec:Rd}, the behavior of some of the $\Gamma_{\rm s}=1.35$ models is to strongly contract upon losing their outermost percent of mass. In these cases,  mass loss is initially stable until these layers are removed. Both the low-density outer envelopes and the initial stability of  mass loss for some models would extend the simulation time beyond what is feasible if these systems were initialized at the same separation as the $\Gamma_{\rm s}=5/3$ donor models. We therefore initialize these systems at smaller $a_0$, from which point they proceed toward coalescence. 

\subsection{Model Evolution and Results}

Snapshots of each of the model systems at decreasing binary separation are shown in Appendix \ref{sec:appendix} in Figures \ref{fig:qslice} through \ref{fig:strucslice}. These may be compared to highlight differences in binary mass ratio (models E, A, F in Figure \ref{fig:qslice}), donor synchronization (models A--D in Figure \ref{fig:syncslice}), gas adiabatic index (models G--I in Figure \ref{fig:adsliceq01} and models  J--L in Figure \ref{fig:adsliceq03}), and donor structure (models I and M in Figure \ref{fig:strucslice}). 

In each case, as time progresses the binary separation tightens over time. Mass is removed from the donor star and expelled from the binary system in the vicinity of the accretor.  Across the models, we observe that there a one-to-one relationship between mass loss from the donor and mass loss from the binary. There is a time delay in that the instantaneous mass loss rate from the donor is reflected at later times in the mass loss rate from the binary; it takes material some time to transit the binary and be lost.  We measure the gas mass in the Roche lobe volume around the accretor and have found that it is small, and does not grow even as the mass loss rate from the donor increases as a function of time (for model F with $q=0.3$, the gas mass enclosed in the accretor Roche lobe is never more than $5\times 10^{-3} M_1$, or less than 1.7\% of the accretor mass). These statements imply that our model results are applicable to phases of rapid mass transfer in which negligible accumulation within the accretor's Roche lobe occurs.

The intensity and morphology of the outflow from the binary depend on the binary separation (as has been discussed in more detail by \citet{2018ApJ...863....5M} and \citet{2018ApJ...868..136M}) and the properties of the binary. A comparison of the flow properties visualized in Appendix \ref{sec:appendix} to the derived properties of the model binary systems  is very useful, and we reference these figures in the sections that follow.


\section{Donor Mass Loss Rate}\label{sec:donormdot}
Next, we analyze how the mass loss rate from the donor varies with time and binary separation in our simulation models. Because this mass loss rate sets the normalization of the overall orbital evolution rate, as seen in equation \eqref{orbit_analytic}, understanding the details of mass loss and exchange is crucial for understanding a binary's orbital dynamics. We compare the mass loss rates to the analytic estimate of \citet{1972AcA....22...73P}, given in equation \eqref{mdot_analytic}.

\subsection{Donor Star Radii: Rotation and Mass Loss}\label{sec:Rd}

The instantaneous mass loss rate from the donor is dictated in part by the degree to which the donor overflows its Roche lobe, as seen in the analytic mass loss rate expression of equation \eqref{mdot_analytic}. In a coalescing binary, the tightening binary separation modifies the Roche lobe radius and degree of Roche lobe overflow as a function of time. Additionally, rotation and mass loss modify the donor's volume-averaged radius, as we discuss in the following subsections. 

\subsubsection{Rotating Simulation Initial Conditions}
The  stars in our simulated systems are both represented on the three-dimensional computational mesh (which has necessarily limited resolution), and put into solid-body rotation.  We relax the initial conditions onto the computational domain (as described in Section \ref{sec:method} and \citet{2018ApJ...863....5M}), then measure the numerically-achieved volume-averaged radius of the donor star. We use a model snapshot from immediately after the relaxation period, prior to the introduction of the gravity of the accretor object. In this snapshot, we compute the volume-averaged radius from the sum of the volume of zones of density greater than a threshold density, $\rho_{\rm thresh} =10^{-4} M_1 / R_1^3$, as
\beq
R_V = \left[  \frac{3}{4\pi} \sum_{\rho > \rho_{\rm thresh}} dV  \right]^{1/3}.
\eeq
In general, as the rotation rate increases stars become more oblate and have increased volume-averaged radii. The tabulated values of the volume averaged initial radii are included in Table \ref{simtable}. 

\subsubsection{1D Adiabatic Mass--Radius Relations}

As the donor loses mass, its structure changes and it settles into a sequence of new hydrostatic equilibria.  We therefore need to know the mass--radius relation of our particular donor models in order to characterize how their radius changes in response to mass loss. We compute the adiabatic response of our donor models to mass loss \citep{1987ApJ...318..794H}. Adiabatic perturbations are relevant because there is no cooling included in our hydrodynamic models. To construct adiabatic variants of our initial donor star model, we follow the method outlined in \citet{2010ApJ...717..724G}. We integrate the expression of hydrostatic equilibrium, $dP/dr = -g \rho$, where $g=Gm/r^2$ and $m$ is the enclosed mass. However, we fix $\rho$ based on an algebraic relationship such that the the specific entropy as a function of mass is identical to the original model.  Given a model profile,  we compute the initial profile of  the polytropic constant, $K_0 = P_0 / {\rho_0}^{\gamma_{\rm ad}}$. Then, within the adiabatic variant model, $K(m) = K_0(m)$, such that $\rho(m) = \left[ P/K_0(m) \right]^{1/\gamma_{\rm ad}}$.

\begin{figure}[tb]
\begin{center}
\includegraphics[width=0.48\textwidth]{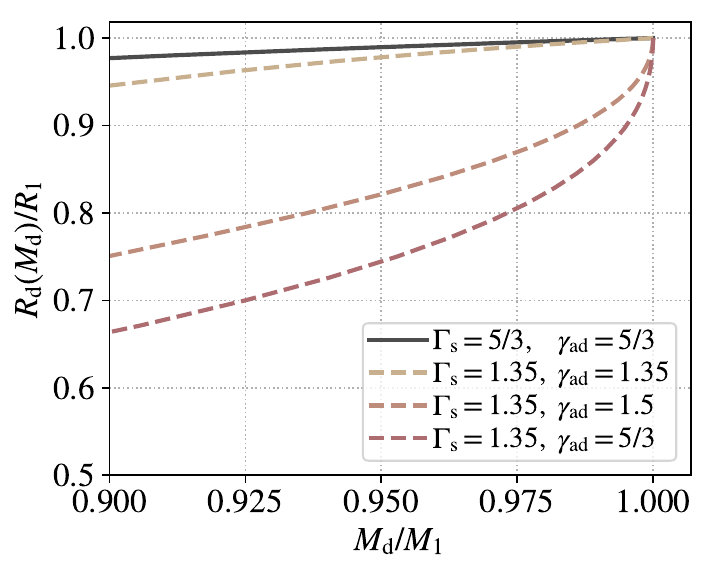}
\caption{1D hydrostatic mass--radius relations of our donor-star models under adiabatic variations. Each of our donor models contracts somewhat as it loses mass, but the response depends on the donor's original entropy profile and the adiabatic index of the equation of state. }
\label{fig:donorMR}
\end{center}
\end{figure}

Figure \ref{fig:donorMR} presents mass--radius relations for our donor models in hydrostatic equilibrium subject to adiabatic mass loss.  Each of the donor models contracts under the influence of mass loss, but to widely varying degrees. The models with $\gamma_{\rm ad} > \Gamma_{\rm s}$ contract most intensely at the onset of mass loss because their highest-entropy outer layers are removed \citep{1987ApJ...318..794H,2010ApJ...717..724G}.

\subsection{Donor Mass Loss Rates and Comparison to Analytic Theory}

We are now in a position to derive mass loss rates from our donor stars in our hydrodynamic simulations and compare them directly to the predictions of the analytic theory, equation \eqref{mdot_analytic}. 

\subsubsection{$\dot \Md$ as a Function of Binary Mass and Separation}

We measure the donor's mass $\Md$ in our simulation models by recording the mass within the original donor radius, $R_1$. To derive $\dot \Md$, we take the time derivative of the tabulated donor mass as a function of time. We find that instantaneous values of the donor mass loss rate are highly variable on timescales less than the donor-star dynamical time as turbulence in the mass-exchange region impacts the instantaneous transfer rate. To focus on the overall dynamical evolution, we will analyze smoothed mass loss rates, with smoothing over 1, 3, or 10 $t_{\rm dyn,1}$ for separations of less than 75\% of the Roche limit, 75 to 90\% of the Roche limit, and greater than 90\% of the Roche limit separation, respectively. 

\begin{figure}[tb]
\begin{center}
\includegraphics[width=0.48\textwidth]{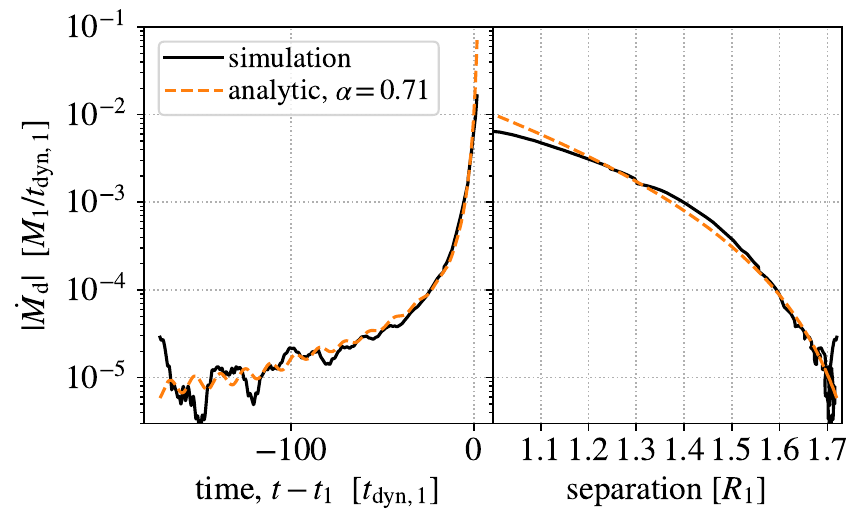}
\caption{ Mass loss rate from the donor versus time and separation. As the simulation progresses and the binary separation decreases, the donor's mass loss rate increases by orders of magnitude. The overall trend is reproduced very well by the analytic expression, equation \eqref{mdot_analytic}, shown here with a best-fit normalizing constant of $\alpha=0.71$.  }
\label{fig:mdot_time_sep}
\end{center}
\end{figure}

Figure \ref{fig:mdot_time_sep} shows an example of the donor's mass loss rate in our fiducial simulation model A (Table \ref{simtable}). Viewed as a function of time, the donor's mass loss rate begins at a small value, but it dramatically increases as the time of merger approaches (here $t_1$ is the time when the separation is equal to $R_1$). Over the duration of this particular simulation, the mass loss rate varies by three orders of magnitude. Viewed in units of binary separation, there is a more progressive trend. As the binary separation decreases, the donor's mass loss rate increases, leading to runaway orbital decay \citep{2018ApJ...863....5M}. 

Figure \ref{fig:mdot_time_sep} also shows equation \eqref{mdot_analytic} for $n=1.5$, which corresponds to $\Gamma_{\rm s}=5/3$ \citep{1972AcA....22...73P} We derive a best-fit value of the normalizing constant $\alpha$, which we denote $\langle \alpha \rangle$, by least-squares minimization of the difference between $\log_{10} \dot \Md$ measured from the simulation and predicted by equation \eqref{mdot_analytic}. In evaluating the degree of Roche lobe overflow in equation \eqref{mdot_analytic}, we use the adiabatic donor star radius as modified by rotation and mass loss, as described in Section \ref{sec:Rd}. In the case of simulation A, we derive $\langle \alpha \rangle = 0.71$, which sets the normalization of the analytic line plotted in Figure \ref{fig:mdot_time_sep}.  When correctly normalized and supplied with time-varying donor radii, the analytic expression captures the overall trends and dependencies of $\dot \Md$ in both time and separation to a remarkable degree of accuracy.

\subsubsection{$\dot \Md$ and Binary Parameters}

\begin{figure*}[tb]
\begin{center}
\includegraphics[width=0.99\textwidth]{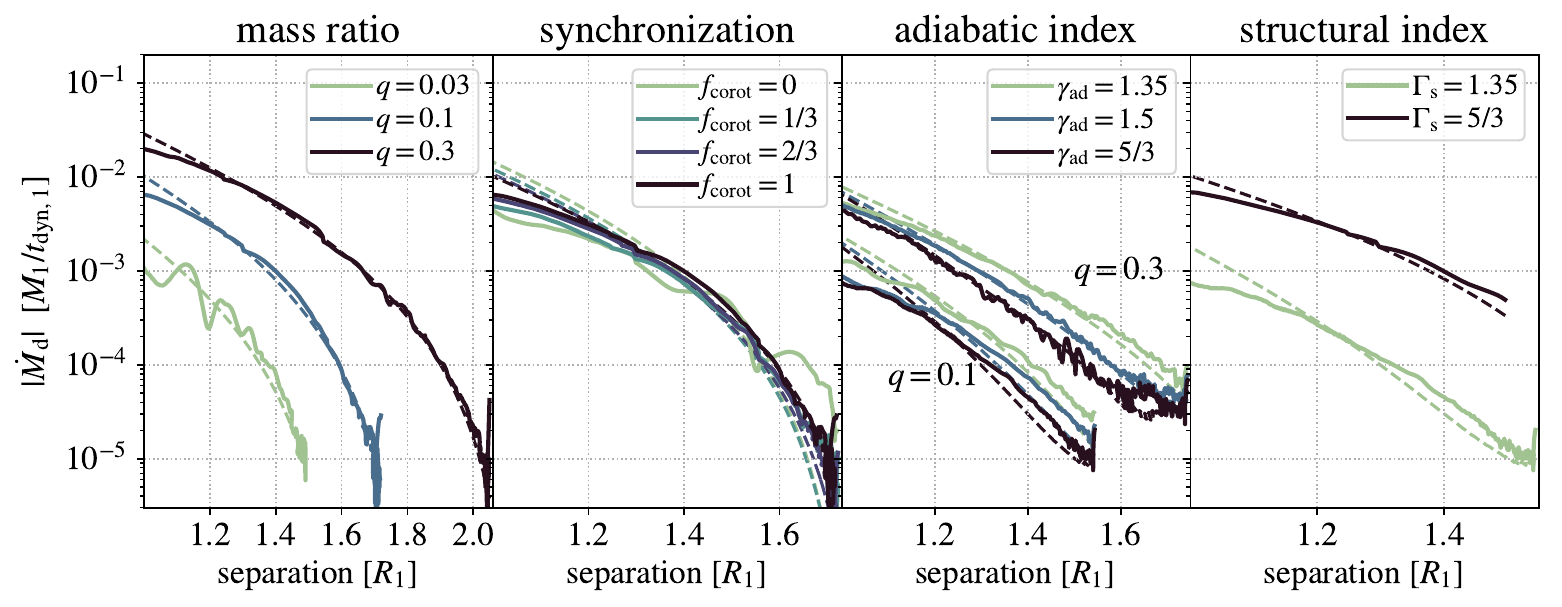}
\caption{ Mass loss rate from the donor star, $\dot \Md$, as a function of binary separation, $a$, in units of the original donor radius. Each panel shows a set of simulations that vary a given binary parameter. Solid lines show the simulation results, while dashed lines of the same color show the predictions of equation \eqref{mdot_analytic} with the best-fit normalization. While varying degree of synchronization and gas adiabatic index each have a small effect on   $\dot \Md$, changing mass ratio or donor structure each can lead to order of magnitude differences in $\dot \Md$ at a given orbital separation.  Comparison to the dashed lines shows that the analytic expression of equation \eqref{mdot_analytic} captures the major trends and shape of $\dot \Md$ with separation.  The mass ratio panel shows models E,A, and F, the synchronization panel shows A--D, the adiabatic index panel shows G--I and J--L, and the structural index panel shows M and I.  }
\label{fig:mdot_variations}
\end{center}
\end{figure*}

In Figure \ref{fig:mdot_variations} we explore the dependence of donor mass loss rate on binary system properties including the mass ratio, degree of donor synchronization, gas adiabatic index, and donor structural index. Solid lines show the simulation results.  For each model, we also plot the analytic prescription, normalized by the best fit coefficient, $\langle \alpha \rangle$, for that simulation. 

Mass ratio, $q$, plays a large role in $\dot \Md$ as a function of binary separation. For larger mass ratio, we observe substantially higher $\dot \Md$ at a given separation. However, this trend is successfully matched by the analytic model, equation \eqref{mdot_analytic}, as shown by the dashed lines.  This indicates that the degree of Roche lobe overflow, $\Rd - R_{\rm L}$, is the primary difference between these sequences. The mass ratio affects the size of the donor's Roche lobe at a given separation, so that when $q$ is larger, the degree of Roche lobe overflow is also higher. 

Donor structural index plays a similarly large role in determining $\dot \Md$ as a function of binary separation. The structural index panel of Figure \ref{fig:mdot_variations} shows models I and M, with $q=0.1$, $\gamma_{\rm ad}=5/3$, and differing $\Gamma_{\rm s}$. When $\Gamma_{\rm s}=5/3$, the donor mass loss rate is about an order of magnitude higher than when $\Gamma_{\rm s}=1.35$ at a given separation. Again, we find that the analytic model reproduces the main features of this dependence. Here the difference lies in the concentration of mass as a function of radius within the donor model. When $\Gamma_{\rm s}=5/3$ the donor is less centrally-condensed, and there is more mass at large radii. In terms of the analytic model, this enters through the exponent on the degree of Roche lobe overflow in equation \eqref{mdot_analytic}. 

We observe much more minimal dependencies of the donor mass loss rate on donor synchronization state or the adiabatic index of our ideal-gas equation of state. In these cases, the primary features that depart from the analytic prediction are time-dependent oscillations of $\dot \Md$ having to do with tidal excitation of waves in the donor star \citep{2019ApJ...877...28M}, and seen in model snapshots in the Appendix.

\subsubsection{The value of $\langle \alpha \rangle$}

We have demonstrated that variations in $\dot \Md$ can be significant under changes to certain properties of the binary system, but in each case, our results are approximated well by the analytic formula. Here we report on our measurements of the best-fit normalizing constant, $\langle \alpha \rangle$, for each model.

\begin{figure*}[tb]
\begin{center}
\includegraphics[width=0.99\textwidth]{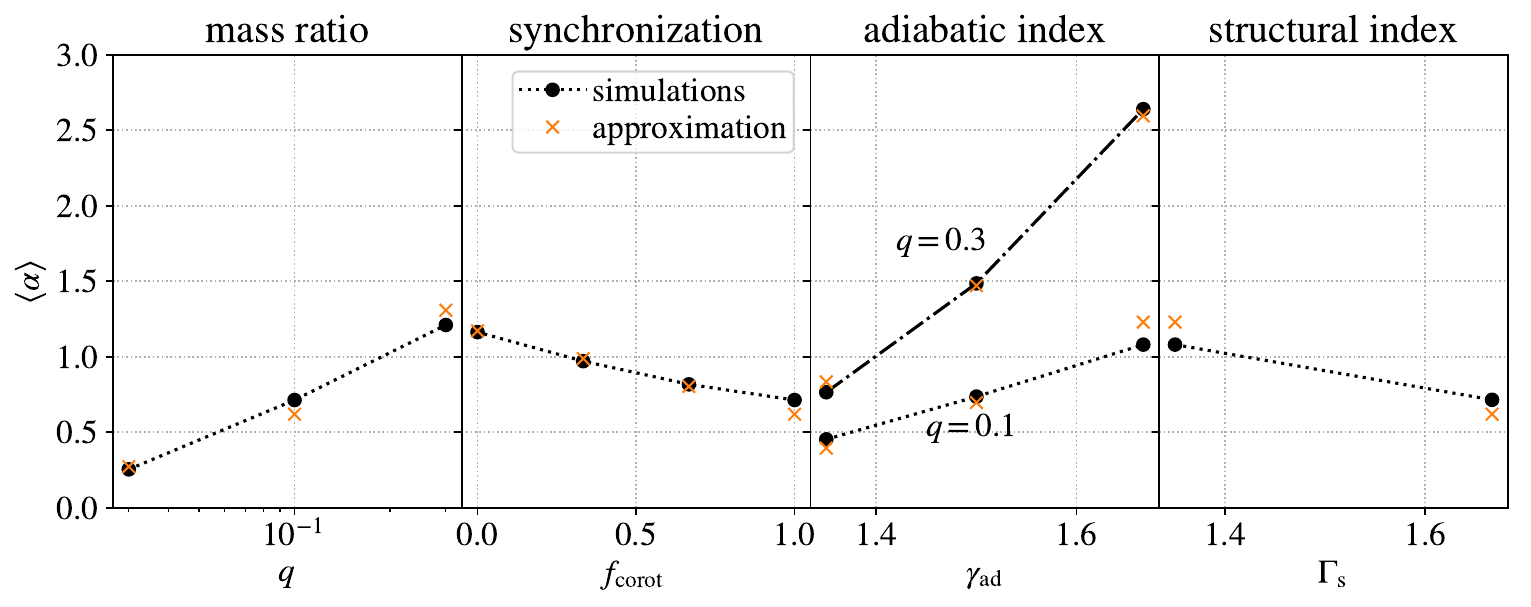}
\caption{ Time-averaged normalization, $\langle \alpha\rangle$, of the analytic mass-loss rate expression, equation \eqref{mdot_analytic}, as measured from the simulations. Typical values of $\langle \alpha\rangle$ are of order unity, but we observe systematic variations with binary parameter variations, particularly with mass ratio and gas adiabatic index. Points labeled approximation derive from the fitting formula of equation \eqref{fit_alpha}.   }
\label{fig:mdot_alpha_variations}
\end{center}
\end{figure*}

Figure \ref{fig:mdot_alpha_variations} shows  values of $\langle \alpha \rangle$ from our derivation under the binary parameter variations. We find values of $\langle \alpha \rangle$ that are distributed within a factor of a few of unity for the cases we have modeled. Thus, $\langle \alpha \rangle \sim 1$ seems to largely approximate the behavior of our data. However, we do find some systematic variations in $\langle \alpha \rangle$ with binary parameters, which we also explore in Figure \ref{fig:mdot_alpha_variations}. In particular, $\langle \alpha \rangle$ increases with $q$ and $\gamma_{\rm ad}$, but decreases with $f_{\rm corot}$ and $\Gamma_{\rm s}$.  We use least-squares minimization to fit an approximating form of the variation of $\langle \alpha \rangle$ with model parameters as
\begin{align}\label{fit_alpha}
\alpha \approx& 0.62 \left( \frac{q}{0.1} \right)^{0.68}  \left( \frac{\gamma_{\rm ad}}{5/3} \right)^{5.39}  \left( \frac{\Gamma_{\rm s}}{5/3} \right)^{-3.25}  \nonumber \\
& \times \left[ 1 - 0.89 \left(f_{\rm corot}-1 \right) \right].
\end{align}
 We caution that while the fit parameters are shown to multiple digits the key uncertainty is not in the least-squares fit but in the fidelity of the underlying simulations (and their extrapolation to astrophysical binary systems). 
In Figure \ref{fig:mdot_alpha_variations}, the over-plotted points labeled approximation come from this function, and reproduce the main trends of the data.  We note that a caveat to equation \eqref{fit_alpha} is that because the parameters are largely varied independently in our experiments, we cannot assure that this functional form applies outside the parameter combinations that have been tested.

It is interesting to speculate on the possible interpretation of some of the parameter dependencies of equation \eqref{fit_alpha}.  Some of the strong dependence on $\gamma_{\rm ad}$ can be traced to the derivation of equation \eqref{mdot_analytic}. Equation \eqref{mdot_analytic} originates from the idea that $\dot \Md = (\rho v)_{L_1} S$, where $S$ is a cross sectional area of the mass-transfer stream. By relating the degree of Roche lobe overflow to the density from the stellar model and the cross sectional area from the potential, we arrive at the proportionality of equation \eqref{mdot_analytic}. It is also typically derived assuming that $\Gamma_{\rm s}= \gamma_{\rm ad}$. When rederived following appendix A of \citet{1972AcA....22...73P} and Chapter 7.1 of O. Pols' notes (see footnote 1) for  $\Gamma_{\rm s}\neq \gamma_{\rm ad}$, we find a prefactor of $\gamma_{\rm ad}^n$.  For $n=2.85$ ($\Gamma_{\rm s}=1.35$), this would account for some of the observed scaling, but not the full exponent of 5.39.  
Next, we observe that the coefficient on $q$ is similar to 2/3. In the derivation of \eqref{mdot_analytic}, it is assumed that only the degree of Roche lobe overflow $(\Rd - R_L)/\Rd$ affects the stream cross sectional area. If the stream width depends on the hill radius of the accretor, with proportionality $q^{1/3}$, we would arrive at a scaling $S\propto q^{2/3}$. Looking particularly at the $L_1$ stream in the left-hand panels of Figure \ref{fig:qslice}, we note that as $q\rightarrow 1$, the equipotential contours and  mass loss stream show broader width.

\section{Specific Angular Momentum of Expelled Material}\label{sec:gloss}

\begin{figure*}[tb]
\begin{center}
\includegraphics[width=0.99\textwidth]{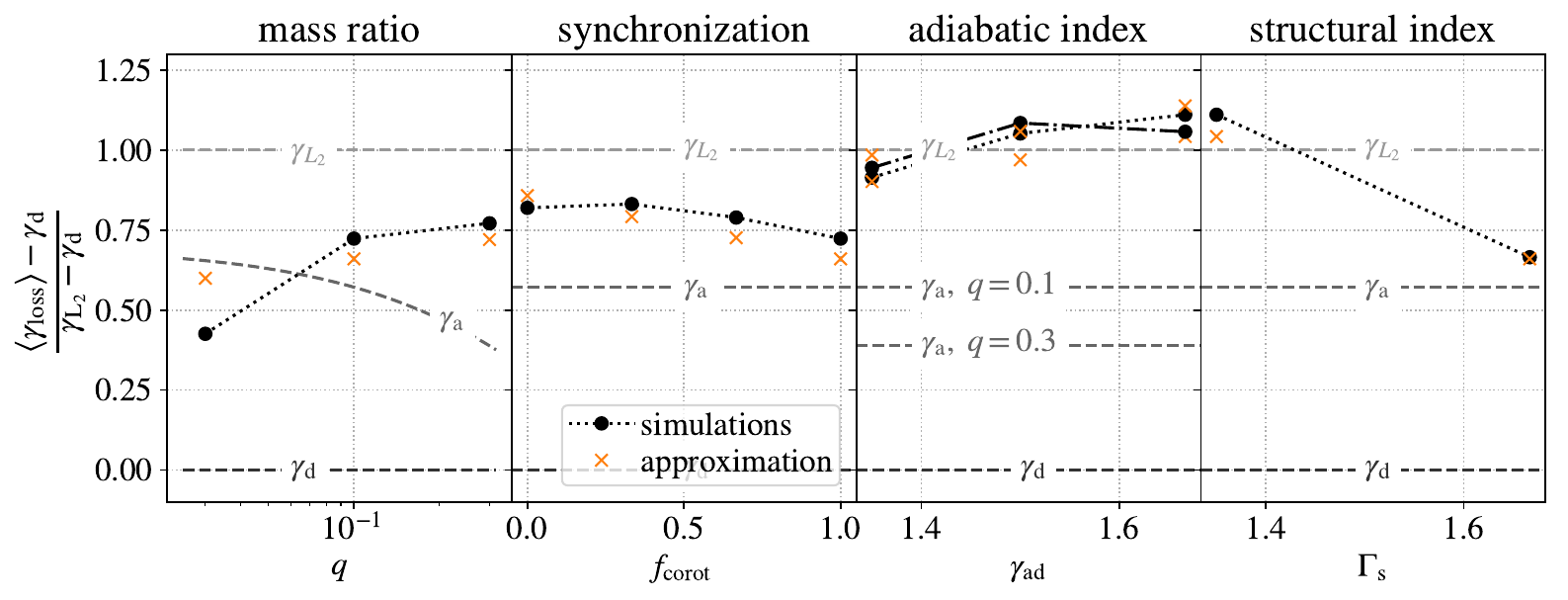}
\caption{ Time averaged specific angular momentum $\langle \gloss \rangle$ carried away from the binary with mass loss, and its dependence on binary properties. We normalize the results in this figure and the approximating form of equation \eqref{fit_gamma} to the range of specific angular momenta from the donor star to the $L_2$ Lagrange point. For most binary parameters, $\gloss$ is intermediate between $\gacc$ and $\gL$. For models with $\Gamma_{\rm s}=1.35$, we find $\gloss \sim \gL$.    }
\label{fig:gamma_loss_variations}
\end{center}
\end{figure*}

Another key parameter that is a priori unknown in the semi-analytic model is the specific angular momentum that is carried away from the binary with mass loss. We report the values of the specific angular momentum using the dimensionless form, $\gloss$. Rather than reporting an instantaneous value, we find that it is useful to measure to perform a mass-loss-weighted average of $\gloss$ over the separation interval from $a_0$ to $R_1$, which we denote with brackets: $\langle \gloss \rangle$. We note that a separation, rather than mass, weighted average differs by at most 15\% from the mass-weighted results. 

In Section \ref{sec:analytic}, we introduced the characteristic values to which $\gloss$ may be compared, the specific angular momentum of the donor, accretor, and that of the outer Lagrange point, $L_2$.  In Figure \ref{fig:gamma_loss_variations}, we present the values of $\langle \gloss \rangle$ with varying binary system parameters. We plot our results on a normalized scale, $(\langle\gloss\rangle - \gd) / (\gL -\gd)$, such that values range from zero (representing the donor's specific angular momentum) to one (representing the angular momentum of $L_2$). This normalization is helpful, because otherwise the primary trend is the dependence of  each of the values of $\gamma$ on $q$. 

We find that in nearly all of our simulations $\langle \gloss \rangle$ lies between the specific angular momentum of the accretor and that of the outer Lagrange point. Qualitatively this fits with the morphology of  a binary system overflowing its Roche lobe. Gas is pulled from the donor (in the process acquiring specific angular momentum at the expense of the accretor's orbital motion) and expelled from the vicinity of the accretor, preferentially toward $L_2$. The geometry of the outflow near $L_2$ varies with the depth of Roche lobe overflow, and the degree of synchronization of the donor \citep{2018ApJ...863....5M,2018ApJ...868..136M}, introducing some of the variations we observe in $\langle \gloss \rangle$. 

Figure \ref{fig:gamma_loss_variations} shows that $\langle \gloss \rangle$ is not particularly sensitive to the initial degree of synchronization or to the gas adiabatic index. It does, however, appear to depend on mass ratio, $q$. Higher mass ratio systems, with more massive accretors relative to the donor, impart higher normalized specific angular momentum to the mass lost.  We additionally find that the structural index of the donor's envelope makes a relatively large difference in $\langle \gloss \rangle$. Donors with $\Gamma_{\rm s} = 1.35$ (shown in the adiabatic index and structural index panels) have higher $\langle \gloss \rangle$ than donors with $\Gamma_{\rm s} = 5/3$. The difference in these models is the degree of central concentration of the donor-star's structure: the lower $\Gamma_{\rm s}$ models have much more tenuous outer layers. Examining model snapshots, we suggest that this difference may be related to the donor star's adiabatic response to mass loss. The $\Gamma_{\rm s} = 1.35$ donor star contracts significantly, and the accretor remains ``skimming" further above the surface, rather than plunging into the donor interior for the separations ($a\geq R_1$) for which we measure $\langle \gloss \rangle$. This in turn affects the gas dynamics of the mass-ejection region. More material is expelled by the $\Gamma_{\rm s} = 5/3$ models because the same total angular momentum is more broadly redistributed across mass. 

We derive an approximating formula for $\langle \gloss \rangle$, that reproduces the main trends of our simulation. We adopt the same functional form as equation \eqref{fit_alpha}, and again use least-squares minimization to fit the parameters. We find, 
\begin{align}\label{fit_gamma}
\frac{\langle \gloss \rangle - \gd}{\gL -\gd} \approx& 0.66 \left( \frac{q}{0.1} \right)^{0.08}  \left( \frac{\gamma_{\rm ad}}{5/3} \right)^{0.69}  \left( \frac{\Gamma_{\rm s}}{5/3} \right)^{-2.17}  \nonumber \\
& \times \left[ 1 - 0.30 \left(f_{\rm corot}-1 \right) \right].
\end{align}
 As with equation \eqref{fit_alpha}, we note that the purpose of this expression is to encapsulate the current simulation results. We do not have sufficient information to comment on the robustness of these models with respect to further physical processes that are not currently modeled. The results of this formula are plotted in Figure \ref{fig:gamma_loss_variations}, labeled ``approximation".

\section{Orbital Decay Dynamics}\label{sec:orb}

In this section, we discuss the dependence of orbit evolution rate on binary model parameters. We apply our measurements of the parameters $\langle \alpha \rangle$ and $\langle \gloss \rangle$ to the point mass orbit evolution expression, equation \eqref{orbit_analytic}, and compare the results of these integrations to the full simulation models. 

\subsection{Dependence on Binary Parameters}

\begin{figure*}[tb]
\begin{center}
\includegraphics[width=0.99\textwidth]{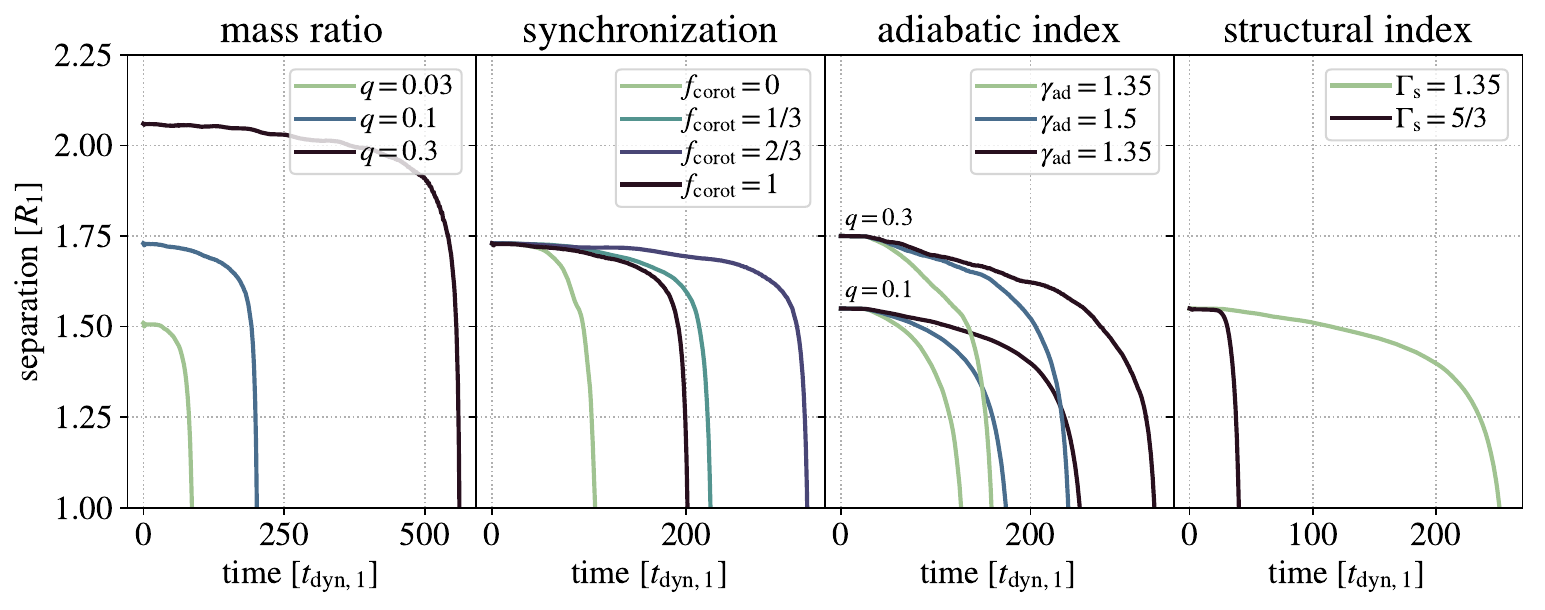}
\includegraphics[width=0.99\textwidth]{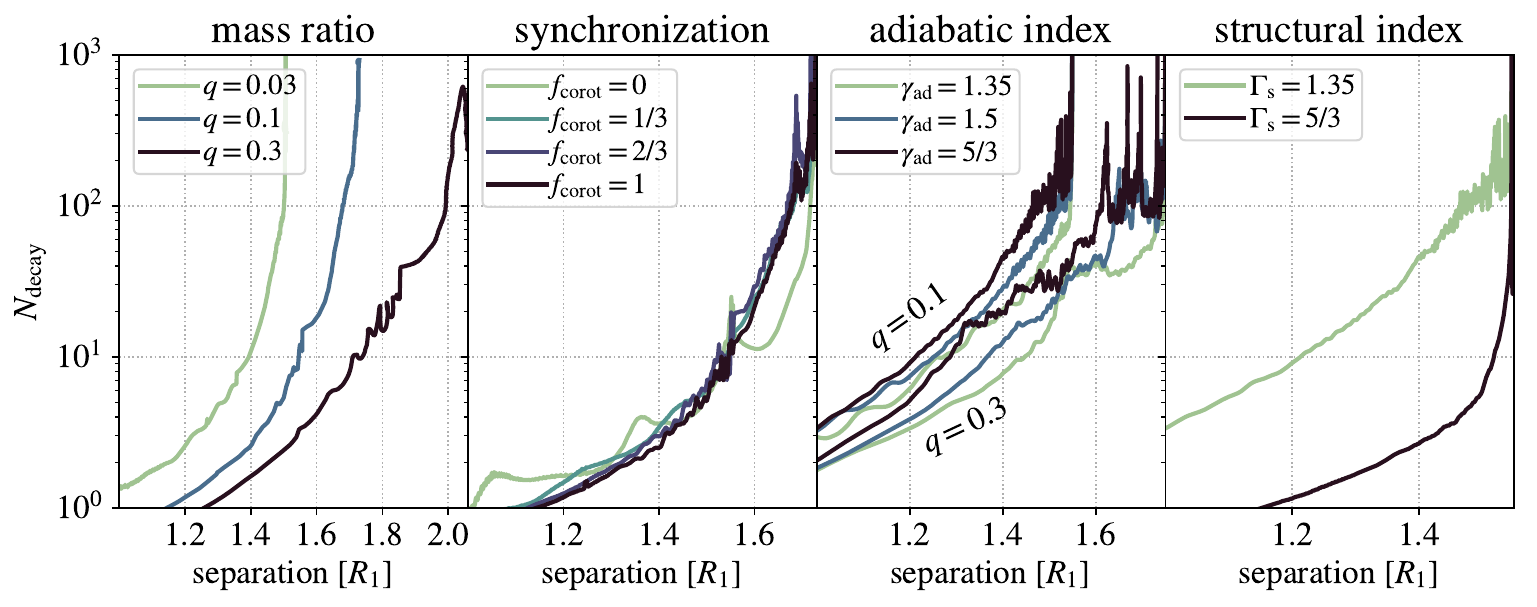}
\caption{ Orbital separation time evolution given varying binary parameters (top panels) and $N_{\rm decay} = \dot \tau^{-1}$  (bottom panels). The rate of orbital decay, as demonstrated by the timescale over which the orbit decreases in separation or $N_{\rm decay}$, is sensitive to the binary mass ratio and the donor star structure, but significantly less sensitive to donor synchronization or gas adiabatic index.    }
\label{fig:orb_variations}
\end{center}
\end{figure*}

Figure \ref{fig:orb_variations} shows two significant aspects of the simulated binary system's orbital evolution. The upper panel shows binary separation as a function of time, while the lower panel shows orbital decay rate as a function of separation. 
The upper panel of Figure \ref{fig:orb_variations} shows orbital separation as a function of time for each model system. While all the models undergo accelerating orbital decay that causes them to plunge toward coalescence \citep{2018ApJ...863....5M}, the relative rates of decay, and thus total simulation durations, are distinct. These differences are reflected in the lower panels of Figure \ref{fig:orb_variations}, in which we present the orbital decay rate, as measured by $N_{\rm decay} = \dot \tau^{-1}$, the inverse rate of change of the orbital period. It therefore represents the number of orbits for the orbit to decay at a given separation. At large separations $N_{\rm decay}$ can be noisy because the  mass loss rate and thus orbital decay rate are slightly variable (we apply the same smoothing as to the donor mass loss rate, described in Section \ref{sec:donormdot}). 

Binary mass ratio leads to obvious differences in the nature of orbital decay. The starting separations for simulations E, A, and F are different -- each is initialized near its particular Roche limit separation.  Overall, the $q=0.3$ model takes much longer to come to the point of coalescence than the $q=0.03$ model. In so doing, the $q=0.3$ model must remove significantly more mass from the donor star in order to be drawn inward. However, examining the lower panels, we see the reflection of the  mass loss rate versus separation shown in Figure \ref{fig:mdot_variations}; at a given separation, the orbit is decaying much faster (over fewer orbits, giving smaller $N_{\rm decay}$) when $q$ is larger. The reason for this can be traced to the higher $\dot \Md$ at a given separation, which we argued is attributable to the larger degree of Roche lobe overflow when the mass ratio is closer to unity. 

By contrast, the degree of synchronization does not have a systematic signature in the orbital separation's decay or in $N_{\rm decay}$. We observe that the different models all coalesce at distinct times, but their variation is not systematic with $f_{\rm corot}$. Instead we find that individual time-dependent variations (especially interactions with tidal oscillations) introduce some stochasticity. For separations less than about $1.6R_1$ we see that  $N_{\rm decay}$ is quite similar for values of $f_{\rm corot}$. 

The value of the adiabatic index, $\gamma_{\rm ad}$, does imprint a systematic difference in the orbital decay properties. In particular, we see that the lower $\gamma_{\rm ad}$, more compressible equations of state lead to somewhat faster orbital decay. Here again, the models with $q=0.3$ have slightly lower $N_{\rm decay}$ at a given separation than those with $q=0.1$. From Figure \ref{fig:mdot_variations}, we see that the mass loss rate from the donor is higher for lower $\gamma_{\rm ad}$, which yields the faster orbital decay (Figure \ref{fig:gamma_loss_variations} shows that the specific angular momentum with which gas is lost is similar across these particular models). The origin of this enhancement in  mass loss rate is the mass-radius relationship of the donor given different $\gamma_{\rm ad}$. The higher $\gamma_{\rm ad}$ models initially contract significantly upon adiabatically losing mass. This reduction of the donor radius decreases the degree of Roche lobe overflow, and consequently $\dot \Md$ and the rate of orbital decay.

Perhaps the most significant distinction in orbital decay properties is seen in the donor-star structure panel on the right-hand side of Figure \ref{fig:orb_variations}. The orbit of the $\Gamma_{\rm s}$ model decays much more rapidly; at a given separation $N_{\rm decay}$ is nearly an order of magnitude larger for $\Gamma_{\rm s}=1.35$ as opposed to $\Gamma_{\rm s}=5/3$.  This difference in normalization is again traceable to the donor's mass loss rate, as shown in Figure \ref{fig:mdot_variations}. The mass loss rate from the $\Gamma_{\rm s}=5/3$ donor is much higher at a given binary separation than that of the $\Gamma_{\rm s}=1.35$ donor. This reflects the difference in the donor star's structures. The $\Gamma_{\rm s}=5/3$ donor star is much-less centrally condensed and therefore has a higher density in its outer layers. An additional, significant factor is the donor mass radius relation for $\Gamma_{\rm s}=1.35$ with $\gamma_{\rm ad}=5/3$, which contracts significantly, also reducing $\dot \Md$ and increasing $N_{\rm decay}$.

\subsection{Reconstruction with Analytic Theory}

Having analyzed the origin of runaway orbital decay in escalating mass loss from coalescing binary systems, we now discuss the reconstruction of binary orbits by integrating the point-mass orbit evolution ordinary differential equations, particularly with parameters motivated by our simulation results. 

\subsubsection{{\tt RLOF} Integration Software}\label{sec:RLOF}

To facilitate orbit modeling with the calibrated point-mass evolution equation, we release a python package {\tt RLOF}\footnote{https://github.com/morganemacleod/RLOF} that solves the initial value problem for the coupled donor mass loss and orbit evolution. The variables that are evolved are the mass and radius of the donor star, $M_{\rm d}$ and $R_{\rm d}$, the mass of the accretor, $M_{\rm a}$, and the separation, $a$. The evolution of these variables is defined by the three equations:
\begin{align}
\dot a &=  -2 a \frac{\dot \Md}{\Md} \nonumber \\
& \times \left[ 1- \beta \frac{\Md}{\Ma}   - \left(1-\beta \right)\left( \gloss + {1\over 2}\right) {\Md \over M}  \right],  \\
\dot \Md & = -  \alpha   {\Md \over \tau } \left( {R_{\rm d} - R_L \over R_{\rm d}} \right)^{n+{3\over 2}}, \\
\dot \Ma & = - \beta \dot \Md. 
\end{align}
These expressions are similar to \eqref{orbit_analytic} and \eqref{mdot_analytic}, with the exception that the accretor is free to accrete a fraction $\beta$ of material lost from the donor. The parameters $\gloss$ and $\alpha$ are set by the approximating equations for $\langle \gloss \rangle$ and $\langle \alpha \rangle$, \eqref{fit_gamma} and \eqref{fit_alpha}, respectively. The donor radius is algebraically defined on the basis of a function $R_{\rm d}( \Md)$, which reflects the adiabatic mass-radius relation.  
Finally, the radius of the Roche lobe $R_{\rm L}$ is set via the \citet{1983ApJ...268..368E} approximation. We note that because $\dot \Md$ increases as $a$ decreases, the solution generally represents superexponetial decay of the orbital separation. 

For comparison with our simulation results, we will take the accretor mass to be constant, which implies that the  mass loss is entirely non-conservative and $\beta=0$. However, in {\tt RLOF}, we also include the option for non-zero $\beta$ either fixed to a constant or limited by the Eddington-limit mass accretion rate of the accretor.

\subsubsection{Results}

\begin{figure}[tb]
\begin{center}
\includegraphics[width=0.4\textwidth]{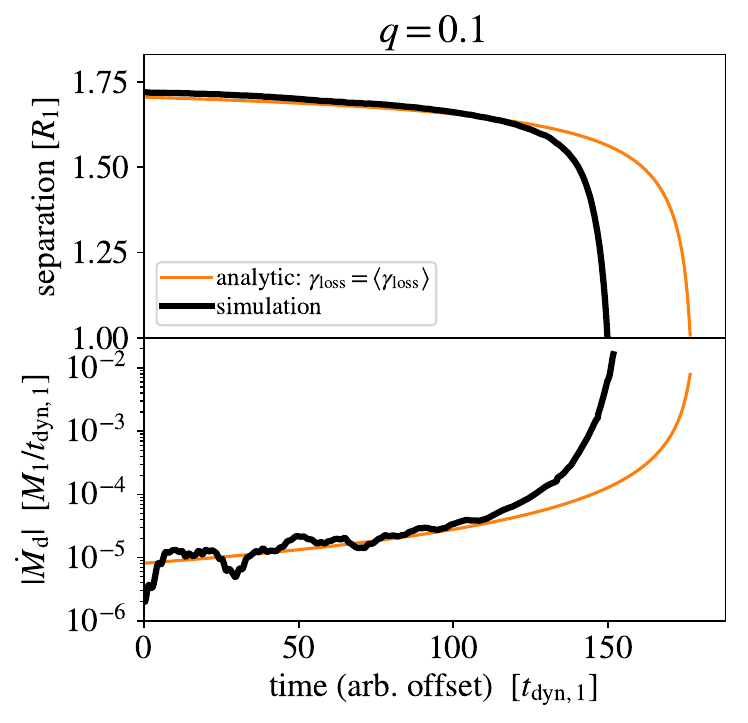}
\includegraphics[width=0.4\textwidth]{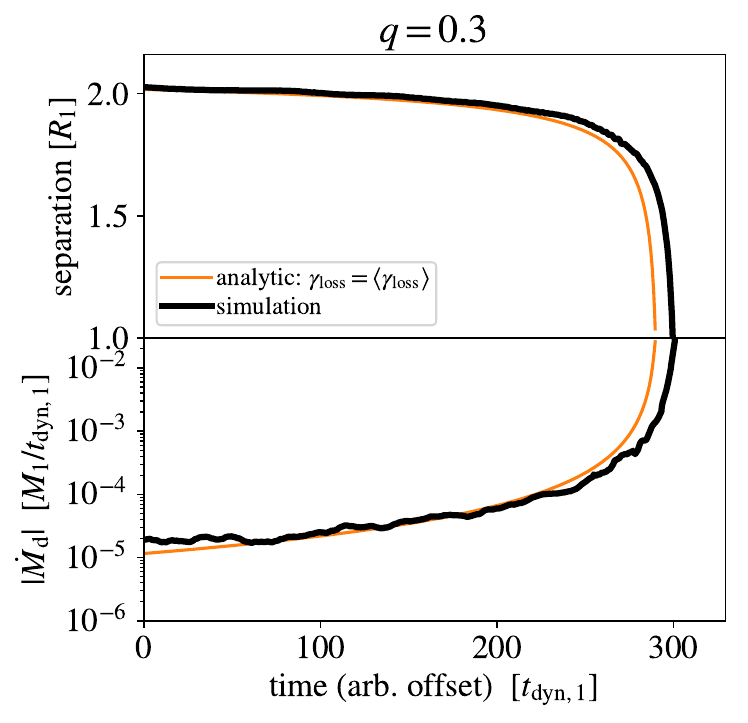}
\caption{ Simulated orbital decay (top panels) and donor mass loss rate (bottom panels) for models A and F, which have mass ratio $q=0.1$ and $q=0.3$, respectively. To compare to reconstructed orbital evolutions with {\tt RLOF}, we initialize the integration with the separation and donor mass and radius 150$t_{\rm dyn,1}$ ($q=0.1$) or 300$t_{\rm dyn,1}$ ($q=0.3$) before $t_1$. We then integrate forward from these initial values given $\alpha=\langle \alpha \rangle$ and $\gloss=\langle \gloss \rangle$, $\gacc$, or $\gL$. The {\tt RLOF} model captures the overall trend of runaway  mass loss and decaying separation to remarkable fidelity. It diverges slightly from the simulated version in cases of variable $\dot \Md$, which lead to compounding differences between the simulation and the {\tt RLOF} model. This effect is most obvious when comparing the more-variable $\dot \Md$ and separation as a function of time for $q=0.1$ to the smoother evolution when $q=0.3$.  }
\label{fig:recon}
\end{center}
\end{figure}

We begin by presenting example orbital evolutions, both as modeled by our hydrodynamic simulations, and as reconstructed by the point-mass integrations of {\tt RLOF}. Figure \ref{fig:recon} shows simulations with $q=0.1$ and $q=0.3$, each initialized synchronously rotating, with $\Gamma_{\rm s} = \gamma_{\rm ad} = 5/3$ (Models A and F). We initialize our {\tt RLOF} integration at a starting time (150 or 300 dynamical times before $t_1$, respectively), at which point we match the binary separation and donor mass. The donor radius is set by the rotating donor radius  and the mass-radius relationship derived in Section \ref{sec:Rd}. We set $ \alpha$ and $\gloss$ using the simulation approximations of equation \eqref{fit_alpha} and \eqref{fit_gamma}. 

Given the runaway nature of orbital decay, small differences along the model's evolution propagate into major distinctions in time of coalescence. For example, variability above or below the predicted $\dot \Md$ creates lasting changes in the binary separation, and in turn, the subsequent evolution. Nonetheless, and even by this very sensitive metric, the example orbital evolutions of Figure \ref{fig:recon} quite reasonably reproduce the overall features of coupled orbital tightening and accelerating donor mass loss. The fit in the case of the $q=0.3$ model is particularly good, while the $q=0.1$ model shows some variability in $\dot \Md$ that propagates into slightly offset time of coalescence. 

The majority of models are reproduced with similar degrees of success by the {\tt RLOF}  model. Some small differences arise. For example, the $\gamma_{\rm ad}$ variations for $\Gamma_{\rm s} = 1.35$ each show that $\dot \Md$ is under-predicted by the analytic expression at $a\gtrsim 1.25 R_1$, by a factor of about two  (Figure \ref{fig:mdot_variations}). This offset in $\dot \Md$ leads to slightly faster coalescence relative to the analytic model (visually similar to the $q=0.1$ case of Figure \ref{fig:recon}). However, these differences are small compared to the differences between respective simulations,  like differences caused by donor-star properties. 

\begin{figure}[tb]
\begin{center}
\includegraphics[width=0.4\textwidth]{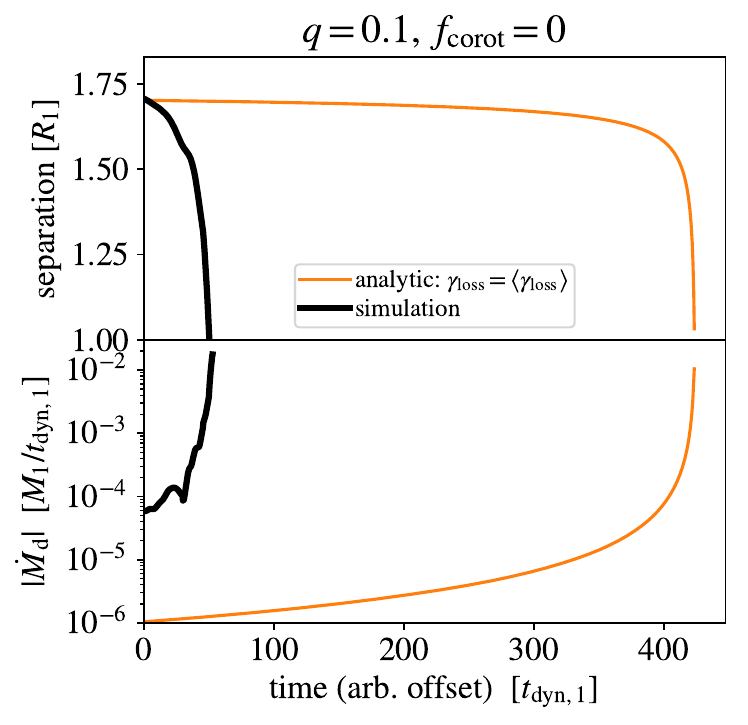}
\includegraphics[width=0.4\textwidth]{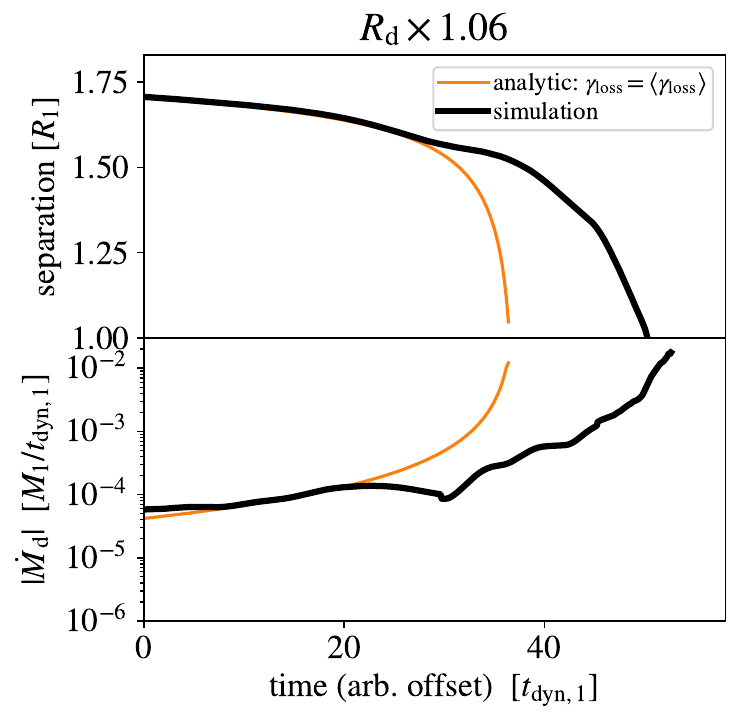}
\caption{ Reconstructed orbital evolution with {\tt RLOF} for the worst-approximated case, with $q=0.1$ and $f_{\rm corot}=0$ case (Model D).  The upper panel shows that matching starting conditions yields a wildly different orbital evolution in which the simulation transfers mass nearly two orders of magnitude more rapidly than predicted and the orbit decays much more rapidly. Inspection of the simulation in question shows very strong resonantly excited tidal waves on the donor's surface. These have the effect of easing  mass loss from the wave peaks, and suppressing it in the wave troughs. With the exception of this time-dependent behavior, the overall rate of decay is much better reconstructed when the donor is artificially made 6\% larger in the {\tt RLOF} integration to approximate the enhanced effective donor size due to tidal oscillations.  }
\label{fig:recon_fc0}
\end{center}
\end{figure}

The most problematic case is the model with $q=0.1$ and $f_{\rm corot}=0$. We show this evolution in Figure \ref{fig:recon_fc0}.  In this case, the simulation $\dot \Md$ is nearly two orders of magnitude higher than predicted at the the outset. This is due to resonant tidal waves (with amplitudes of approximately 10\% the donor's radius) that set up on the donor star and more-easily allow material to escape the donor's Roche lobe, see Figure \ref{fig:syncslice} and \citet{2019ApJ...877...28M}. The lower panel of Figure \ref{fig:recon_fc0} reproduces the {\tt RLOF} integration with the donor radius artifically enhanced by 6\%, more correctly capturing the overall mass loss rate, but not modeling the time-dependent features in $\dot \Md$ that have to do with individual tidal wave peaks and troughs interacting with the accretor \citep{2019ApJ...877...28M}. 

\section{Discussion}\label{sec:discussion}

In this section we discuss the application of our simulation results, through the {\tt RLOF} model, to observations of binary systems decaying toward coalescence. 

\subsection{V1309 Sco}

The binary system V1309 Sco merged in 2008, accompanied by a 10 magnitude optical flare \citep{2010A&A...516A.108M}. Later \citet{2011A&A...528A.114T} identified the progenitor source as an eclipsing binary with a decreasing orbital period in the years prior to outburst.  In the months before the dramatic outburst, periodicity disappeared and the luminosity gradually rose. \citet{2014ApJ...788...22P} demonstrated that non-conservative mass loss from a binary could explain the period decrease and eventual optical enshroudment. Later \citet{2017ApJ...850...59P} undertook more detailed modeling of the light curve morphology to based on ray tracing through smoothed particle hydrodynamic models of mass loss from the $L_2$ point of a binary system. 

Given the abundance of information available on V1309 Sco, this is an ideal object on which to test modeling of the binary coalescence process. It is clear from V1309 Sco's pre-outburst lightcurve that the object's photosphere initially traces the two Roche lobes of the binary system \citep{2011A&A...528A.114T}. This evidence has been interpreted as indicative of a contact system in which both stars fill their Roche lobes in hydrostatic equilibrium \citep[e.g.][]{2011A&A...528A.114T,2014ApJ...788...22P,2017ApJ...850...59P} or of a non-conservatively mass transferring system in which material pulled from the donor fills, then overfills, the accretor's Roche lobe \citep{2014ApJ...786...39N}.  Either case might have similar pre-outburst light curve morphology because gas fills both the donor and accretor Roche lobes. 

\begin{figure}[tb]
\begin{center}
\includegraphics[width=0.48\textwidth]{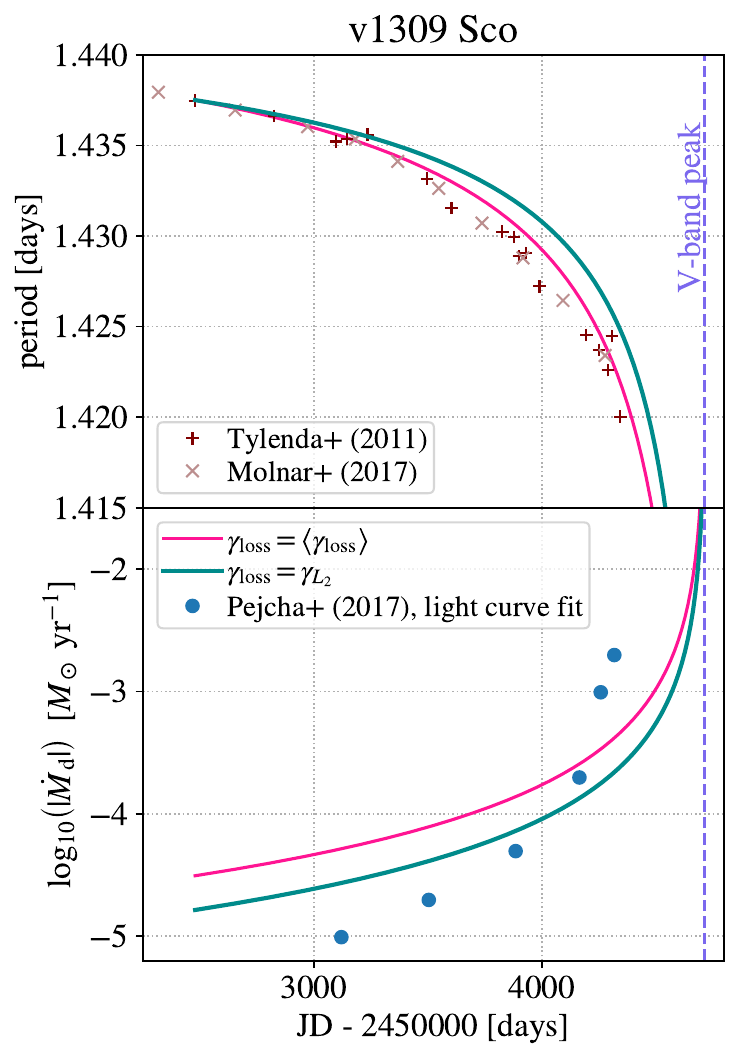}
\caption{ Time evolution of the orbital period and donor mass loss rate for V1309 Sco from data and {\tt RLOF} modeling. {\tt RLOF} models are initialized to match the first orbital period reported in \citet{2011A&A...528A.114T}, and to reach a separation of the donor's original radius at the time of the outburst's V-band peak.  Thus constrained, we are able to compare the relative fit of models with differing specific angular momenta of material lost.  In the donor mass loss rate panel, we also compare to estimates from the lightcurve derived by \citet{2017ApJ...850...59P}, as described in the text. }
\label{fig:v1309}
\end{center}
\end{figure}

 We explore the non-conservative mass loss model further through comparison to our hydrodynamic simulations. 
Rather than attempting to perform an identical hydrodynamic simulation, here we apply the {\tt RLOF} model, which facilitates some exploration of the possible system parameter space. We show two example {\tt RLOF} integrations in Figure \ref{fig:v1309}. The upper panel of Figure \ref{fig:v1309} compares the orbital period decrease to measurements tabulated by \citet{2011A&A...528A.114T}  and \citet{2017ApJ...840....1M}. The lower panel shows the corresponding mass loss rate from the donor, and compares to the estimates of \citet{2017ApJ...850...59P}, which are derived by fitting radiative transfer models with a given mass loss rate to the lightcurve shape.

The two model lines in Figure \ref{fig:v1309} represent the different assumptions about the specific angular momentum of material carried away from the binary:
 \begin{enumerate}[label=(\roman*)]
\item  $\gloss = \langle \gloss \rangle$,   which is based on simulation results for our models of mass loss from the donor and binary; and 
\item $\gloss = \gL$,  which would likely be relevant to an overcontact binary system in hydrostatic equilibrium in the rotating reference frame \citep{2014ApJ...788...22P,2017ApJ...850...59P,2019MNRAS.489..891H}. 
\end{enumerate}
We initialize both models at an orbital period equal to that of the first data point from \citet{2011A&A...528A.114T}, approximately $1.437$~d. For each model, we choose the donor radius, and hence the degree of Roche lobe overflow, such that the binary merges (reaches separation equal to the donor radius) at the time of the optical peak of the V-band lightcurve, September 6, 2008 \citep[reporting on data from the American Association of Variable Star Observers]{2010A&A...516A.108M}. We adopt $q=0.1$, based on estimates from the eclipsing light curve morphology \citep{2014ApJ...788...22P,2016RAA....16...68Z,2017ApJ...850...59P}.   We assume that the donor star is described by $\Gamma_{\rm s} = \gamma_{\rm ad} = 5/3$, which affects $\langle \gloss \rangle$ and $\dot \Md$ as a function of the degree of Roche lobe overflow. Finally, due to lack of constraints, we make the simplifying assumption of no change in donor radius with mass loss.\footnote{  In particular, the core and envelope mass of the donor are unknown and thermal adjustment of the outer layers are likely important, implying departure from an adiabatic mass--radius relation \citep{2015MNRAS.449.4415P}. One could imagine implementing a more informed donor mass--radius relationship to capture some of these properties. }

As the orbital period decreases from approximately 1.44~d to 1.42~d, the mass loss rate from the donor (and consequently from the binary) increases by an order of magnitude. The upper panel of Figure \ref{fig:v1309} shows that the simulation values for $\gloss=\langle\gloss\rangle$ approximate the curvature of the decaying light curve slightly better than the choice of $\gloss=\gL$.  The choice of specific angular momentum affects the normalization of the instantaneous mass loss rate from the binary, as well as the total mass lost to the circumbinary environment. In both cases, the models are broadly consistent with \citet{2017ApJ...850...59P}'s estimates from the light curve morphology alone (see Fig 5 of  \citet{2017ApJ...850...59P} for a similar comparison).  One area of inconsistency is the steep 

While this particular example fully explores only two model parameter choices, it does illustrate the value of rapid modeling with a calibrated tool like our {\tt RLOF} integrator.  There is reason to believe other parameters may be important. \citet{2014ApJ...788...22P} used a fit of the rate of period decay along with growth of the estimated photosphere radius to constrain the donor-star's polytropic structure and found values $\Gamma_{\rm s} \approx 1.12$ ($n=7.83$) and argued that this might represent nearly-isothermal outer layers of the donor. Thus, further and more complete exploration of possible degeneracies is clearly warranted. Because individual model integrations have negligible computational cost, one could imagine exploring the binary parameter space with Markov Chain Monte Carlo to make more robust inferences on the binary properties leading up to coalescence.   

\subsection{$N_{\rm decay}$--$N_{\rm accel}$ plane}

The fact that coupled equations drive the decay of the orbit and mass loss from the donor have constraining implications for the properties of binary systems trending toward coalescence. Because the mass loss rate from the donor sets the orbit evolution rate and the orbit evolution rate sets the rate at which $\dot \Md$ changes (via the changing degree of Roche lobe overflow), there is a link between orbital decay rate and the rate of change of the orbital decay rate. One observationally motivated way to express these properties is through the period derivative $\dot \tau$ and the period second derivative $\ddot \tau$. We define two properties: the number of orbits over which the orbital period is decaying, $N_{\rm decay} = 1 / \dot \tau$, and the number of orbital periods over which the period decay is accelerating, $N_{\rm accel} = |\dot \tau/\ddot \tau|/\tau$. 

\begin{figure}[tb]
\begin{center}
\includegraphics[width=0.48\textwidth]{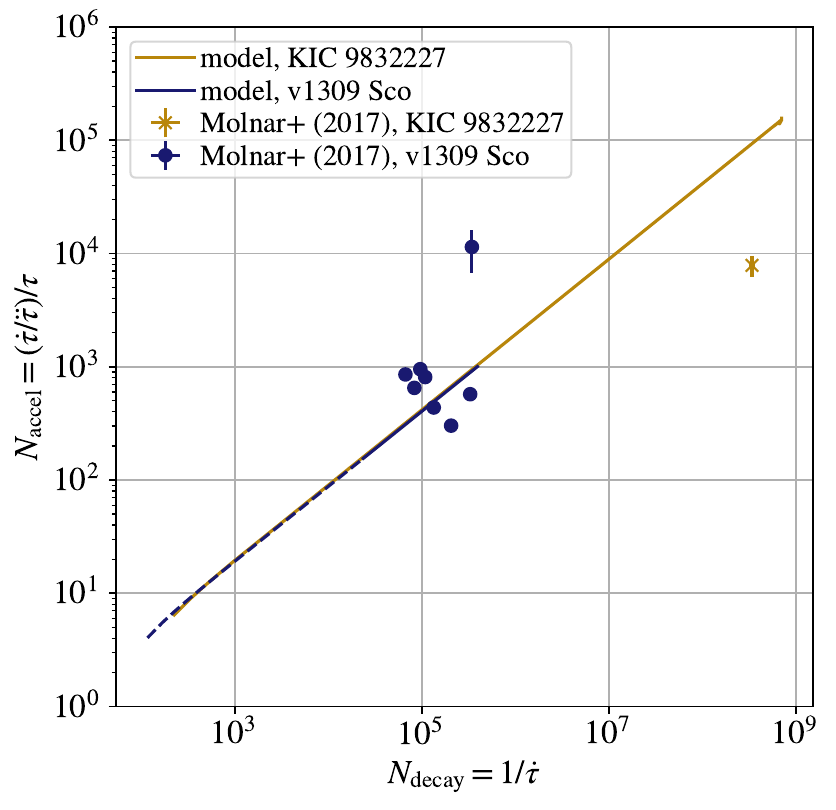}
\caption{ The $N_{\rm decay}$--$N_{\rm accel}$ plane with a well-defined track set by models of stellar orbital decay due to unstable  mass loss. We show {\tt RLOF} models matching the initial period derivatives of V1309 Sco and KIC 9832227. The dashed line shows the period range where V1309 Sco had become obscured \citep{2014ApJ...788...22P}. The overplotted data are from \citep{2017ApJ...840....1M}. Errors are propagated on the KIC 9832227 from reported one sigma errors. For V1309 Sco, no errors are tabulated by \citep{2017ApJ...840....1M}, so errors are propagated assuming similar fractional error to KIC 9832227 (However, the scatter indicates that the errors may be underestimated). The bulk of the data points for V1309 Sco lie along a locus of similar $N_{\rm decay}$--$N_{\rm accel}$ to the {\tt RLOF} model. However, for KIC 9832227, when $N_{\rm decay}$ is consistent with the data, the measured $N_{\rm accel}$ is approximately an order of magnitude too low to be consistent with the {\tt RLOF} model. We find that across parameter variations, the {\tt RLOF} model $N_{\rm decay}$--$N_{\rm accel}$ tracks are well approximated by equation \eqref{NNapprox}. }
\label{fig:NN}
\end{center}
\end{figure}

Figure \ref{fig:NN} shows the $N_{\rm decay}$--$N_{\rm accel}$ plane. We plot two models in this parameter space, one discussed above for V1309 Sco (with $\gloss = \langle \gloss \rangle$), and the other that matches the orbital period, masses, and reported period derivative of KIC 9832227 \citep{2017ApJ...840....1M}. KIC 9832227 was originally reported to be a candidate for a future merger episode. Some of the ephemeris on which this claim was based have since been disproven \citep{2018ApJ...864L..32S}, with a more likely scenario that the system is a hierarchical triple \citep{2019A&A...631A.126K}. However, more important for our present purposes is the model used to extrapolate forward and predict the data of merger. \citet{2017ApJ...840....1M} fit the same functional form as \citet{2011A&A...528A.114T}, an exponential of arbitrary scale length, to their data. However, we see in Figure \ref{fig:NN} that the reported $N_{\rm accel}$ is far too small to be consistent with the reported $N_{\rm decay}$. A binary with a small period derivative should also have a proportionately small period second derivative if driven by runaway, unstable  mass loss. We note that when varying binary parameters over a wide range in mass ratio, donor structure, etc, we always find nearly consistent values, within a half dex of,
\beq\label{NNapprox}
\log_{\rm 10} \left( N_{\rm accel} \right)  \approx {2 \over 3} \log_{\rm 10} \left( N_{\rm decay} \right) -0.7.
\eeq
This implies that the following combination of derivatives is a constant, 
\beq
\frac{\dot \tau^{5/3} }{\ddot \tau \tau} \approx 0.19.
\eeq
The $N_{\rm decay}$--$N_{\rm accel}$ plane can thus serve as a useful discriminant of whether candidate merger progenitors are consistent with the unstable  mass loss scenario. 

\subsection{Limitations and their Importance}\label{sec:limitations}

It is important to acknowledge several limitations of our simulation and orbit-reconstruction methodology that may affect the application of the {\tt RLOF} model  to realistic binaries. The full covariances of binary parameters have not been fully explored. The present primary limitation is that we have chosen to perform our hydrodynamic model calculations with relatively high spatial resolution of the donor star and mass loss region. This is important to preserve the stability of the quasi-hydrostatic donor star, and to accurately model the progression from gradual to rapid  mass loss \citep{2018ApJ...863....5M}. However, this choice leads to a moderate computational expense for each model,\footnote{The expense is on the order of $5\times 10^4$ cpu-hours per model evolution using intel broadwell processors.} which necessarily limits the parameter space that may be explored in methodologically related work \citep[for example, see][]{2014ApJ...786...39N,2018MNRAS.481.3683K}. 

A key simplification that we make as a result of these limitations is  adopting two model donor star structures, as modeled by polytropic envelopes (with polytropic index $\Gamma_{\rm s}$) surrounding a central condensed core. This gives an indication of the range of possible outcomes given different donor star structures, but cannot guarantee that a realistic donor structure will be accurately represented. Similarly, we adopt an ideal gas equation of state, and vary the adiabatic index $\gamma_{\rm ad}$ through the range of possibility. Finally, and perhaps most importantly, we do not co-vary each of the possible parameter combinations and instead largely perform one dimensional surveys along each axis of our selected parameter space.  This greatly reduces the computational expense at the cost of uncertainty on the validity of results in untested parameter combinations. 

In the future, targeted calibration surveys of hydrodynamic models could be performed with, for example, realistic donor stars, equations of state, and more realistic treatment of physical processes that may affect how mass is lost from the binary system. One such physical process is radiative heating and cooling, as this will introduce scale dependence into our presently dimensionless hydrodynamic models. A lower-density, larger radius donor will have different ratio of radiative diffusion timescale to dynamical timescale, and effectively more compressible gas in some parts of the flow. Magnetohydrodynamical stresses are similarly not presently modeled and could be important in removing material from the rotating flow around the accretor \citep[for a parallel discussion of these concerns, see][]{2019MNRAS.489..891H}. 

Despite these limitations to our present approach, our current results do provide guidance as to the degree of sensitivity of the parameters that describe binary orbital evolution to differences that arise across the parameter space of binary systems.

\section{Summary and Conclusions}\label{sec:conclusion}
We reported on a parameter survey of three-dimensional hydrodynamic models of binary system coalescence. These models apply to giant stars merging with more compact companions such as main sequence stars or compact objects (the accretor is unresolved in our simulations and is treated only gravitationally, as a point mass). In each case, we initialized the model system at a separation less than the Roche limit but greater than the donor's initial radius, and observe the phase of unstable  mass loss and orbital tightening that ensues. 

For each of the models we have simulated,  mass loss is, by construction, unstable, in that it runs away to higher rates and drives the orbit to decay toward coalescence. We analyzed the phase during which the separation is greater than the donor's original radius, and considered models with varying mass ratio, donor star spin and structure, and gas equation of state (see Table \ref{simtable}). Rather than realistically pertaining a particular binary system, these models were intended to span some of the possible parameter space of pre-common envelope binary interactions. 

Some of the key findings of this analysis are as follows:
\begin{enumerate}

\item Orbital evolution is driven by  mass loss from the donor and the torques this material imparts on the binary. We measure two key uncertain parameters -- the donor star mass loss rate and the specific angular momentum imparted to material lost from the donor -- using our hydrodynamic simulations. These measurements complete the ability to solve coupled, semi-analytic expressions for donor mass loss and orbit evolution, as described in Section \ref{sec:analytic}. 

\item We find that the functional form of donor star mass loss rates as a function of time and separation are robustly well-matched  by the model of \citet{1972AcA....22...73P}, equation \eqref{mdot_analytic} (Section \ref{sec:donormdot} and Figure \ref{fig:mdot_variations}). We compute the best-fit value of the normalizing constant $\alpha$ for the donor mass loss rate. We find that typical values are similar to one, with variation by a factor of a few. Figure \ref{fig:mdot_alpha_variations} shows how $\alpha$ varies with changing binary model parameters, and we present an approximating formula in equation \eqref{fit_alpha}.

\item We compute the mass-averaged specific angular momentum carried away from the binary by material lost from the donor in Section \ref{sec:gloss}. Typical values of the specific angular momentum of the material removed from the donor then lost from the binary are between the angular momentum of the accretor and the outer, $L_2$ Lagrange point. We examine the ways that $\langle \gloss \rangle$ varies with binary parameters in Figure \ref{fig:gamma_loss_variations}, and present an approximating formula in equation \eqref{fit_gamma}. 

\item In Section \ref{sec:orb}, we discuss the resultant orbital decay and how it differs significantly in binary systems with differing parameters. The orbital decay rate is a function of the donor mass loss rate and is, therefore, particularly sensitive to the binary mass ratio and the donor star structure. Donor models with centrally condensed $\Gamma_{\rm s} = 1.35$ envelopes yielded much slower orbital evolution than donors with $\Gamma_{\rm s}=5/3$ envelopes (Figure \ref{fig:orb_variations}). 

\item We demonstrate that, when calibrated with measurements from simulations, the point mass orbit evolution expressions reproduce the main behaviors of runaway orbital decay and escalating donor mass loss rate. We encapsulate a tool to perform these integrations into a python package, {\tt RLOF}, which is publicly released accompanying this work (Section \ref{sec:RLOF}). We demonstrate the performance and shortcomings of this tool in Figures \ref{fig:recon} and \ref{fig:recon_fc0} using our simulation models. 

\item We demonstrate that the {\tt RLOF} model can reproduce the observed orbital decay and time of outburst of V1309 Sco (Figure \ref{fig:v1309}). We find that orbital decay from unstable Roche lobe overflow leads to a well-defined track in a plane defined by orbital period, and its first and second derivatives. This $N_{\rm decay}$--$N_{\rm accel}$ plane is shown in Figure \ref{fig:NN} and may prove to be a useful metric to identify and validate eclipsing binary candidates for future binary merger. 
\end{enumerate}

In being based on a limited set of hydrodynamic models, there are very likely parameter combinations of astrophysical relevance that are not modeled in this work. Additionally, as we discuss in Section \ref{sec:limitations}, the specific angular momentum imparted to material lost from the binary is sensitive to the gas flow around the accretor. This implies that currently unmodeled physical  properties like radiative diffusion and cooling or the magnetic field of the donor star could conceivably  impart meaningful deviations from the dimensionless gas dynamical models  we consider here. 

Future refinements notwithstanding, the degree to which the {\tt RLOF} point-mass model can reproduce the main features of  binary coalescence is very promising for the application of this rapid integration method to future binary merger candidates, rapid binary population synthesis models, and the construction of initial conditions for hydrodynamic models of later stages of common envelope phases.

\acknowledgements{M.M. gratefully acknowledges the contributions of E.C. Ostriker and J. M. Stone, whose advice and insights were central to the formative stages of this work and, especially, the method it is based on. We additionally thank O. De Marco, M. Mapelli, J. Grindlay, E. Ramirez-Ruiz, and S. Toonen for helpful discussions. Finally, we gratefully acknowledge constructive and thoughtful feedback from the anonymous reviewer. 

We acknowledge support for this work provided by NASA through Einstein Postdoctoral Fellowship
grant number PF6-170169 awarded by the Chandra X-ray
Center, which is operated by the Smithsonian Astrophysical
Observatory for NASA under contract NAS8-03060. 
This material is based upon work supported by the National Science Foundation under Grant No. 1909203. 
Resources supporting this work were provided by the
NASA High-End Computing (HEC) Program through
the NASA Advanced Supercomputing (NAS) Division
at Ames Research Center. }

\software{IPython \citep{PER-GRA:2007}; SciPy \citep{jones_scipy_2001};  NumPy \citep{van2011numpy};  matplotlib \citep{Hunter:2007}; Astropy \citep{2013A&A...558A..33A}; Athena++ (Stone, J.M., \url{https://github.com/PrincetonUniversity/athena-public-version}) RLOF v1.0 \citep{RLOF1.0}, BinaryOrbitEvolution \citep{morgan_macleod_2020_3662958}. }

\bibliographystyle{aasjournal}

\appendix
\section{Simulation Snapshots and Roche Lobe Overflow Morphology}\label{sec:appendix}

In this appendix, we provide model snapshots at various binary separations for each of the simulation models referenced in this paper. These snapshots are taken as slices of density through the orbital plane of the binary system, and are presented in rotated, $x'$--$y'$ coordinate system such that the accretor, $M_2$, always lies along the $+x'$-axis. Length units are in original donor radii, $R_1$, and density is in units of $M_1/R_1^3$. 

\begin{figure*}[tbp]
\begin{center}
\includegraphics[width=0.99\textwidth]{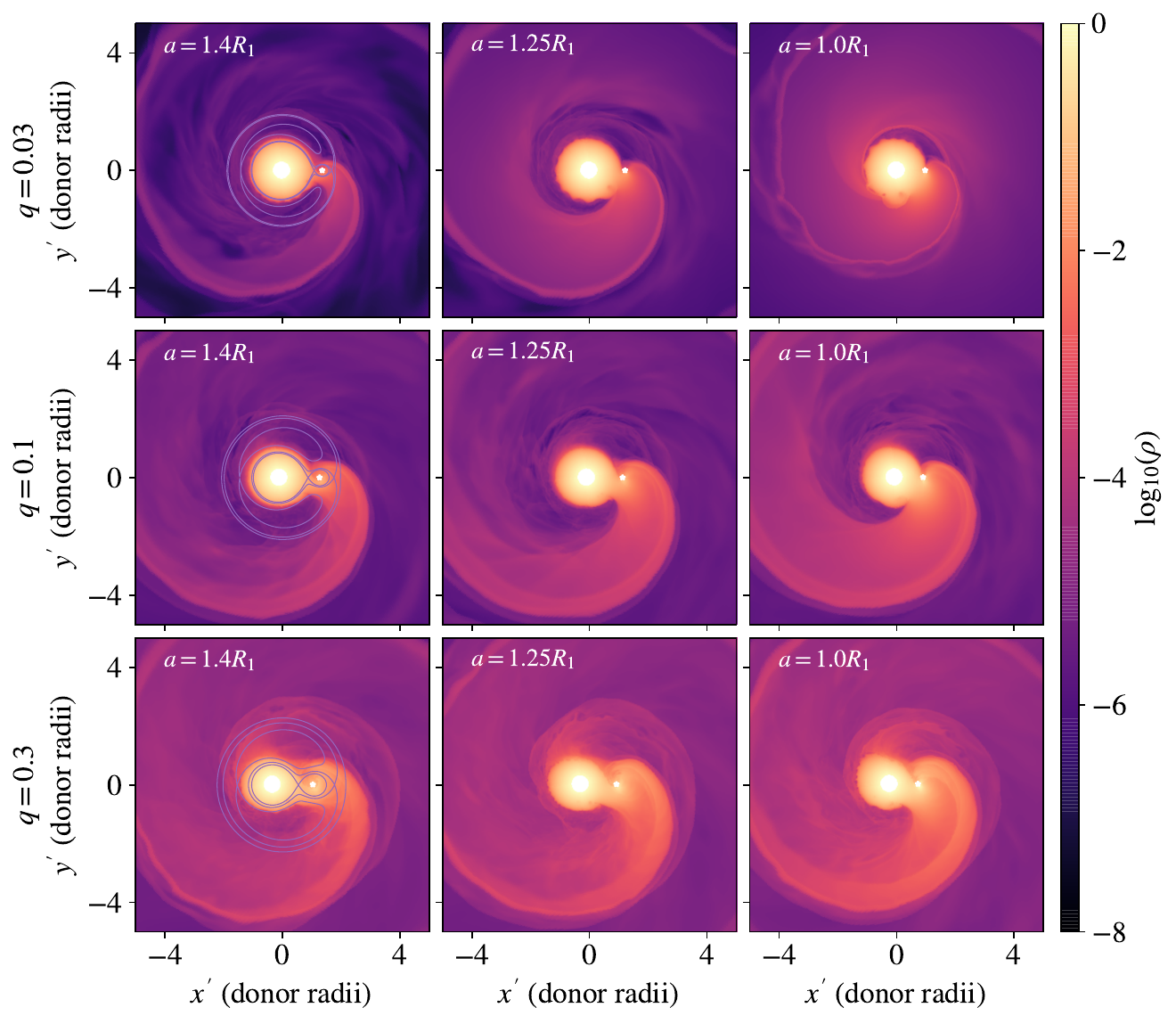}
\caption{Slices through the orbital midplane of binary coalescence models with varying mass ratio. Each row shows a different mass ratio, while columns show fixed binary separations. Slices are plotted in rotated $x'-y'$ coordinates such that the accretor lies along the $+x'$-axis. Density is shown in code units of $M_1/ R_1^3$.  As mass ratio increases we see that the accretor's tidal force distorts the donor star to increasing degrees and that the stream of material flowing away from the binary intensifies. Surface waves excited by resonances between the tidal forcing and the orbital frequency are present in each mass ratio, but are particularly evident in the $q=0.03$ model \citep{2019ApJ...877...28M}.   }
\label{fig:qslice}
\end{center}
\end{figure*}

\begin{figure*}[tbp]
\begin{center}
\includegraphics[width=0.99\textwidth]{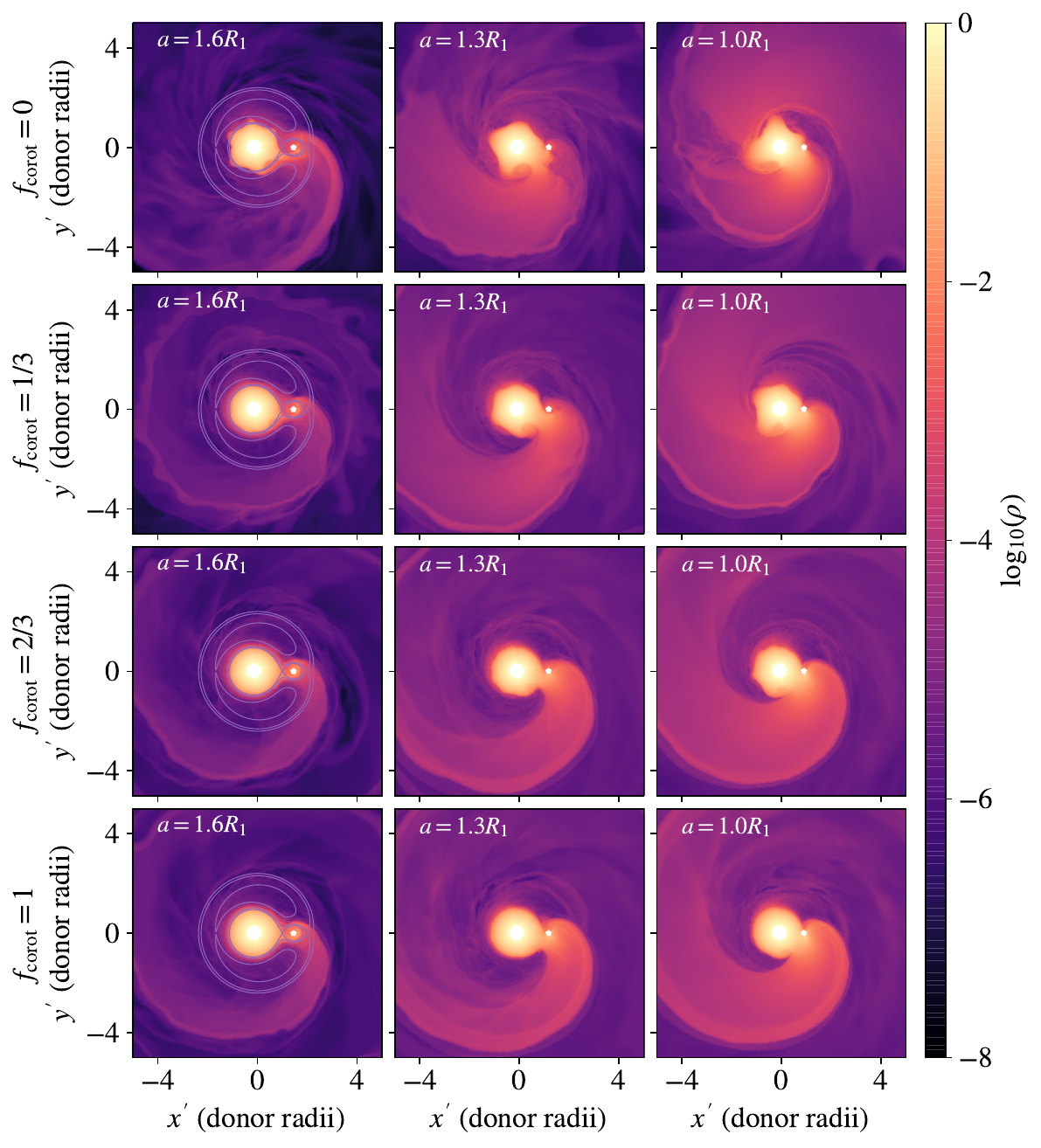}
\caption{Slices of logarithm of gas density through the orbital midplane, as in Figure \ref{fig:qslice}. Each row shows snapshots of models with different initial rotation, parameterized by fraction of corotation angular frequency $f_{\rm corot}$. }
\label{fig:syncslice}
\end{center}
\end{figure*}

\begin{figure*}[tbp]
\begin{center}
\includegraphics[width=0.99\textwidth]{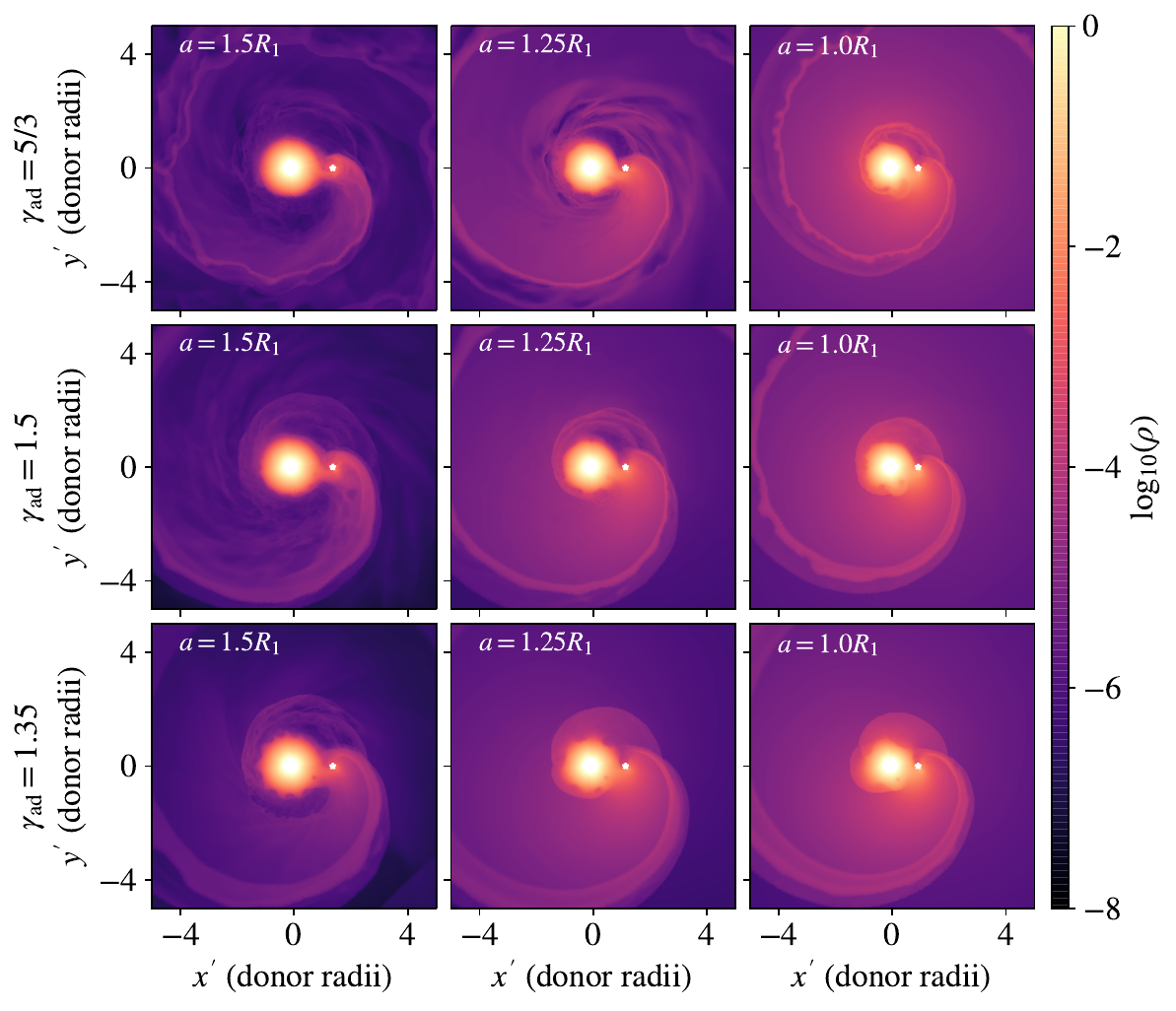}
\caption{Slices of logarithm of gas density through the orbital midplane, as in Figure \ref{fig:qslice}. Each row shows snapshots of models with different adiabatic index, $\gamma_{\rm ad}$. For these models, $q=0.1$. }
\label{fig:adsliceq01}
\end{center}
\end{figure*}

\begin{figure*}[tbp]
\begin{center}
\includegraphics[width=0.99\textwidth]{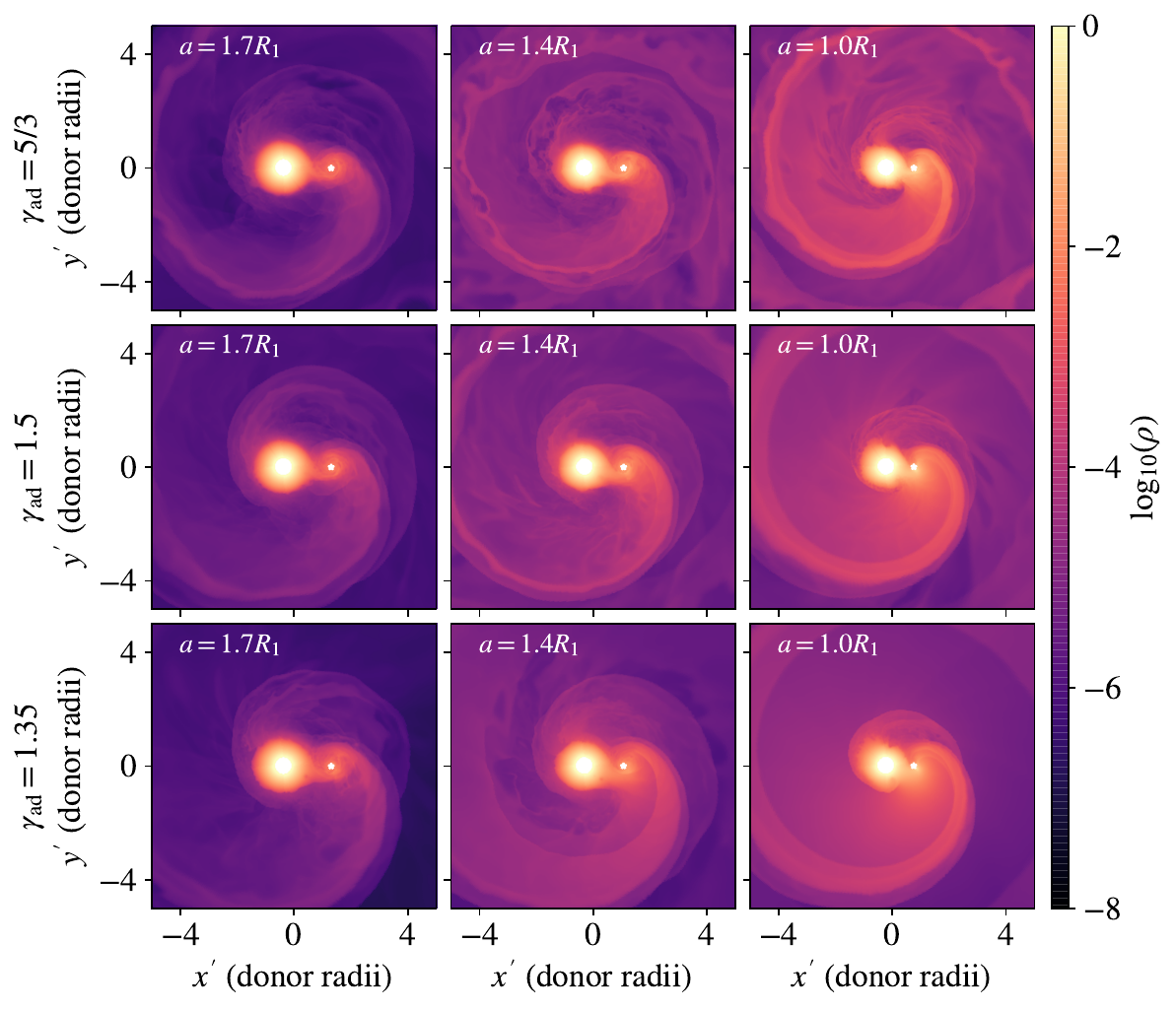}
\caption{Slices of logarithm of gas density through the orbital midplane, as in Figure \ref{fig:qslice}. Each row shows snapshots of models with different adiabatic index, $\gamma_{\rm ad}$. For these models, $q=0.3$. }
\label{fig:adsliceq03}
\end{center}
\end{figure*}

\begin{figure*}[tbp]
\begin{center}
\includegraphics[width=0.9\textwidth]{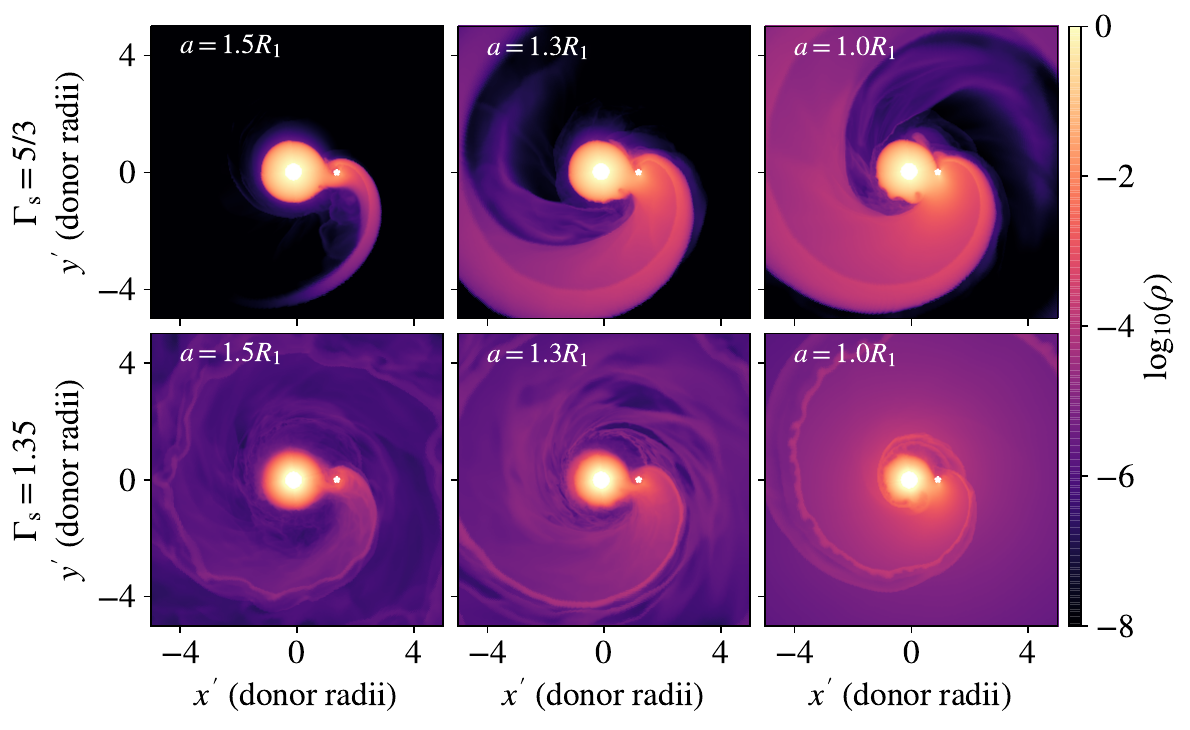}
\caption{Slices of logarithm of gas density through the orbital midplane, as in Figure \ref{fig:qslice}. Each row shows snapshots of models with different donor star structure, parameterized by the polytropic index of the envelope, $\Gamma_{\rm s}$.}
\label{fig:strucslice}
\end{center}
\end{figure*}

The details of the snapshots plotted are as follows:
\begin{itemize}
\item Figure \ref{fig:qslice} shows slices of models with varying binary mass ratio, $q$. From top to bottom row, we show models E, A, and F, respectively. 
\item Figure \ref{fig:syncslice} shows slices of models with varying initial donor rotation, parameterized by $f_{\rm corot}$. From top to bottom row, the models are A, B, C, and D. 
\item Figures \ref{fig:adsliceq01} and \ref{fig:adsliceq03} show models with varying adiabatic index and $q=0.1$ or $q=0.3$, respectively. These models share structural polytropic index $\Gamma_{\rm s}=1.35$. In Figure \ref{fig:adsliceq01}, from top to bottom, the models shown are G, H, and I. In Figure \ref{fig:adsliceq03}, from top to bottom, the models shown are J, K, and L. 
\item Figure \ref{fig:strucslice} shows models with differing structural index, $\Gamma_{\rm s}$. These models are model M (top) and I (bottom). 
\end{itemize}

\section{Software and Data}

Along with the publication of this paper we publicly release software tools and data to reproduce our results and model other binary systems. 

\begin{itemize}
\item The reduced data products of our simulation models, including the data shown in Table \ref{simtable}, the full history of binary motion and integral quantities,  input files associated with each simulation model, and the software to reproduce the figures in this paper are released at \url{https://github.com/morganemacleod/BinaryOrbitEvolution} and \citet{morgan_macleod_2020_3662958}.
\item The {\tt RLOF} software package described in Section \ref{sec:RLOF} is released at \url{https://github.com/morganemacleod/RLOF} and \citet{RLOF1.0}. 
\item The polar-averaging scheme that we employ to reduce the impact of the coordinate singularity at the poles in Athena++ hydrodynamic setups is released along with an example problem at \url{https://github.com/morganemacleod/athena-public-version/tree/polar-zone-avg}
\end{itemize}

\end{document}